\documentclass[a4paper,11pt]{article}
\pdfoutput=1 % if your are submitting a pdflatex (i.e. if you have
\usepackage{jheppub} % for details on the use of the package, please
                     \usepackage{amsmath,amssymb,amsfonts,amsthm}
\usepackage[T1]{fontenc} % if needed
\usepackage{xcolor}
\usepackage{bbm}
\usepackage{booktabs,multirow}
\usepackage{subfigure}
\usepackage{braket}
\usepackage{array}
    \newcolumntype{P}[1]{>{\centering\arraybackslash}p{#1}}
    \newcolumntype{M}[1]{>{\centering\arraybackslash}m{#1}}

\usepackage{hyperref}
\newtheorem{theorem}{Theorem}[section]

\theoremstyle{remark}
\newtheorem{remark}[theorem]{Remark}
\newcommand{\ex}[1]{\mathrm{e}^{#1}}

\newcommand{\pa}[1]{\left(#1 \right)}

\newcommand{\bb}[1]{\mathbb{#1}}

\newcommand{\ca}[1]{\mathcal{#1}}

\newcommand{\abs}[1]{\left|#1\right|}

\newcommand{\kett}[1]{ \ket{#1}\!\rangle }

\newcommand{\ti}[1]{\tilde{#1}}

\newcommand{\fr}{\frac}

\def\be{\begin{equation}}
\def\ee{\end{equation}}
\def\ba{\begin{eqnarray}}
\def\ea{\end{eqnarray}}

 \def\ba{{\bar{\alpha}}}

\def\tr{{\text{tr}}}

\begin{document}

\begin{flushright}
CALT-TH 2023-037
\\
IPMU 23-0030
\\
RIKEN-iTHEMS-Report-23
\\
\end{flushright}

\author[1,2]{Yuya Kusuki}
\author[1,3]{\!\!, Sara Murciano}
\author[1,4]{\!\!, Hirosi Ooguri}
\author[1]{\!\! and Sridip Pal} 
\affiliation[1]{Walter Burke Institute for Theoretical Physics,  California Institute of Technology,  Pasadena, CA 91125, USA}   
\affiliation[2]{ RIKEN Interdisciplinary Theoretical and Mathematical Sciences (iTHEMS), \\Wako, Saitama 351-0198, Japan}
\affiliation[3]{Department of Physics and Institute for Quantum Information and Matter, California Institute of Technology, Pasadena, CA 91125, USA}
\affiliation[4]{Kavli Institute for the Physics and Mathematics of the Universe (WPI), University of Tokyo, Kashiwa 277-8583, Japan}

\title{Symmetry-resolved Entanglement Entropy, Spectra \& Boundary Conformal Field Theory}

\abstract
    {
    We perform a comprehensive analysis of the symmetry-resolved (SR) entanglement entropy (EE) for one single interval in the ground state of a $1+1$D conformal field theory (CFT), that is invariant under an arbitrary finite or compact Lie group, $G$. We utilize the boundary CFT approach to study the total EE, which enables us to find the universal leading order behavior of the SREE and its first correction, which explicitly depends on the irreducible representation under consideration and breaks the equipartition of entanglement. We present two distinct schemes to carry out these computations. The first relies on the evaluation of the charged moments of the reduced density matrix. This involves studying the action of the defect-line, that generates the symmetry, on the boundary states of the theory. This perspective also paves the way for discussing the infeasibility of studying symmetry resolution when an anomalous symmetry is present. The second scheme draws a parallel between the SREE and the partition function of an orbifold CFT. This approach allows for the direct computation of the SREE without the need to use charged moments. From this standpoint, the infeasibility of defining the symmetry-resolved EE for an anomalous symmetry arises from the obstruction to gauging.
   Finally, we derive the symmetry-resolved entanglement spectra for a CFT invariant under a finite symmetry group. We revisit a similar problem for CFT with compact Lie group, explicitly deriving an improved formula for $U(1)$ resolved entanglement spectra. Using the Tauberian formalism, we can estimate the aforementioned EE spectra rigorously by proving an optimal lower and upper bound on the same. In the abelian case, we perform numerical checks on the bound and find perfect agreement. }
\maketitle

\section{Introduction}

Symmetries have a notable impact on the aesthetic appeal of the nature. The presence of symmetries facilitates the formulation of physical laws and conservation principles, leading to profound insights into the behavior of natural phenomena, often described by the framework of statistical and quantum field theories. In recent years, there has been a plethora of activity in expanding the notion of symmetry starting from higher form symmetry \cite{Gaiotto:2014kfa} to non-invertible symmetry, see recent reviews \cite{Cordova:2022ruw,McGreevy:2022oyu}. Besides providing an organizing principle, it presents itself as a tool to extract and constrain the non-perturbative physics \cite{Chang:2018iay,Cordova:2022fhg}. 

The idea of symmetry resolution in the context of evaluating density of states transforming under particular irreducible representation (irrep) of the symmetry appeared in $1+1$D CFT \cite{Pal:2020wwd} and subsequently generalized to higher dimension in the work of \cite{Harlow:2021trr}, leading to the universal prediction for a finite group $G$ with order $|G|$: 
\begin{equation}\label{density}
    \rho_r(E) = \frac{d_r^2}{|G|}\rho(E)\,,\quad E\to \infty \,.
\end{equation}
Here $\rho_r(E)$ is the density of states at energy $E$, restricted to the irrep $r$ while $\rho(E)$ is the full density of states at energy $E$, $d_r$ is the dimension of the irrep $r$. The above equation is verified in the context of free and weakly coupled free theories in \cite{Cao:2021euf} based on \cite{Melia:2020pzd,Henning:2017fpj} and extended to the case of continuous symmetry in \cite{Kang:2022orq} and space-time symmetry in \cite{Mukhametzhanov2020,Benjamin:2023qsc}. Furthermore, in $1+1$D CFT, such a symmetry resolution has been performed for non-invertible symmetries \cite{Lin:2022dhv}, where the notion of topological defect line (TDL) and category theory have played a crucial role. See  \cite{Verlinde:1988sn,Moore:1988qv,Fuchs:2002cm,Frohlich:2004ef,Frohlich:2006ch,Frohlich:2009gb,Carqueville:2012dk,Brunner:2014lua,Aasen:2016dop,Bhardwaj:2017xup,Chang:2018iay,Thorngren:2019iar,Cordova:2019wpi,Lin:2019hks,Pal:2020wwd,Komargodski:2020mxz,Aasen:2020jwb,Thorngren:2021yso,Huang:2021zvu,Kaidi:2022cpf,Lin:2022dhv,Chang:2022hud,Lu:2022ver,Kaidi:2023maf,Zhang:2023wlu,Lin:2023uvm} for various application of TDLs and category theory.

There has been a parallel independent development in the world of statistical physics, where the study of symmetry has intertwined with that of another significant phenomenon i.e.\! entanglement, captured by von Neumann and Renyi entropies. If a theory is endowed with a global symmetry, the interplay between entanglement and the symmetry can be analyzed through a quantity called \textit{symmetry-resolved entanglement entropy} (SREE). Roughly speaking, it can be thought of as the entanglement entropy in a fixed charge sector. One motivation that rose the attention towards this subject was shedding light on the experiment studying the time evolution of the entanglement in systems with many-body localization \cite{Lukin}. This has sparked the interest in developing a theoretical framework to study the symmetry resolution. In particular, an intense activity in the area of 1+1D quantum models started with the works \cite{Goldstein2018,sierra}, some of which predate the development mentioned in the previous paragraph. One of the main results that emerged is the statement that the entanglement entropy is equally distributed among the different sectors of a $U(1)$ symmetric theory, a result known as \textit{equipartition} of the entanglement. Given the relevance of entanglement in the computational cost to simulate quantum many-body systems, this finding is crucial for designing efficient protocols and algorithms. 
A similar feature has also been found in theories with a non-abelian symmetry, like Wess-Zumino-Witten (WZW) models in the continuum case \cite{calabrese2021symmetryresolved,Milekhin2021}.  This feature is also present in studies of finite groups, as demonstrated by studying the $\mathbb{Z}_2$ symmetry of the Ising spin chain \cite{Horv_th_2020,Fraenkel_2020}, the $\mathbb{Z}_3$-symmetric Potts model \cite{Capizzi:2021kys}. Further instances can be found in integrable and free quantum field theories \cite{Murciano_2020,horvath2021u1,horvath2021branch,ghasemi,cdfmssca-22,cdfmssca-22-2,michele}, holographic settings \cite{Zhao_2021,Weisenberger:2021eby,Zhao:2022wnp} and lattice models \cite{Bonsignori_2019,Murciano:2019wdl,Calabrese:2020tci,Tan_2020,Murciano:2020lqq,Bonsignori_2020,Turkeshi_2020,Estienne_2021,Ares:2022hdh,jones-22,pvcc-22,Lau2022,ore-21}. Furthermore, other entanglement measures like negativity \cite{Cornfeld_2018,Murciano_2021,chen2022negativityboson,multicharged,Chen:2021nma,Foligno_2023}, relative entropy \cite{Chen:2021pls}, distances \cite{capizzi2021symmetry}, modular flow \cite{Di_Giulio_2023} have been studied in CFT. Finally, a historical albeit important remark is that a quantity closely related to the charged Renyi, one of the precursors of SREE is defined and explored in the context of holography in\cite{Belin:2013uta} long before all of the above developments. The quantity under consideration in 
 \cite{Belin:2013uta} is not entirely same as the charged Renyi, explored in the community of statistical physics, however, they do share similar physics on a conceptual level and one can be extracted from another. 

One cannot but notice the structural similarity between the resolution of EE and density of states. For example, the symmetry resolution of EE of a single interval of length $\ell$ with respect to $\mathbb{Z}_2$ symmetry of the Ising spin chain reveals
\begin{equation}\label{SREE}
    S_{vN}(\ell,\text{even})=S_{vN}(\ell,\text{odd})=S_{vN}(\ell)+\log(1/2)\,,\quad  \ell/\varepsilon \to \infty
\end{equation}
Here $S_{vN}(\ell)$ is the von Neumann entropy. If we compare eq.~\!\eqref{density} with eq.~\!\eqref{SREE}, it is obvious that $e^{S_{vN}(\ell,r)}$ has a similar behavior as $\rho_r$ while $e^{S_{vN}(\ell)}$ behaves like $\rho$. The striking structural similarity between the study of symmetry-resolved density of states and symmetry-resolved entanglement entropy in the context of $1+1$D CFT naturally leads to a search for a general framework. This is exactly where boundary conformal field theory (BCFT) comes in and plays a crucial role owing to the seminal work \cite{Cardy2004} (see also \cite{Cardy_2016,Calabrese_2010}). In \cite{Cardy2004} the computation of entropy of a single interval was performed by computing an annulus partition function. This has a natural extension in the context of SREE.

In this work, we make the aforementioned analogy precise leveraging the language of BCFT leading to the following dictionary: 

\renewcommand{\arraystretch}{1.5}
\begin{table}[h]
	\centering
	\begin{tabular}{@{}ll@{}ll@{}ll}\toprule
		%\hline
		SRDOS & SR(RE/EE) & \quad SR(RE/EE) in BCFT\\
		\midrule
  Torus Partition function &  Exp[Renyi entropy] & \quad Annulus partition function\\
  Twisted partition function & Charged Moment of Renyi & \quad Twisted partition function\\
   $L/\beta\to \infty $ & $\log(\ell/\varepsilon)\to\infty $ & \quad $\log(\ell/\varepsilon)\to\infty $ \\
		%\hline
  Hamiltonian & Entanglement Hamiltonian& \quad Open string Hamiltonian\\
		Density of states & Density of EE spectra & \quad  Density of states in Open string   \\
		%\hline
		%\midrule
		\bottomrule
	\end{tabular}
	\caption{\label{table:modularestimate} The dictionary between symmetry-resolved density of states and symmetry-resolved entanglement entropy (for single interval) in $1+1$D CFT. SRDOS stands for Symmetry-Resolved Density Of States, SRRE stands for Symmetry-Resolved Renyi Entropy, SREE stands for Symmetry-Resolved Entanglement Entropy. $L$ and $\beta$ are the two periods of the torus, while $\log(\ell/\varepsilon)$ is related to the width of the annulus.}
\end{table}

We remark that the importance of BCFT has been emphasized in \cite{DiGiulio2022} (see also \cite{Northe:2023}), where they performed the computation of SREE with respect to the symmetry groups $U(1)$, $\mathbb{R}$ and $\mathbb{Z}_2$ of $U(1)$ present in the free compact and non-compact boson.\\

The main goal of the present work is to present a general BCFT framework for computing the SREE on one hand while, on the other hand, we bring the tools developed in the context of symmetry resolution of density of states to the setup of SREE. The main results are listed below:

\begin{itemize}
    \item \textit{We prove the universality in the leading order behavior of SREE in any $1+1$D CFT with symmetry $G$, admitting a unique $G$-invariant vacuum and a $G$-invariant boundary state. In particular, we show that for any finite group $G$,
\begin{equation}\label{eq:MAINRESULT}
       \lim_{q\to 1} \left[S_n(q,r)-S_n(q)\right]=\log\frac{d_r^2}{|G|}\,,
    \end{equation}
    where $|G|$ denotes the order of $G$.   Here $S_n(q,r)$ is the $n$-th Renyi entropy within irrep $r$, $S_n(q)$ is the full R\'enyi entropy and $q$ is a parameter fixed by the geometry and the UV cut-off. For example, if we consider an interval of length $\ell$ of a $1+1$D CFT on infinite line, $q=e^{-\pi^2/\log(\ell/\varepsilon)}$. Here $q\to1$ limit amounts to taking the UV cut-off $\varepsilon$ to $0$ relative to the physically relevant length scale (for example, length of a spatial interval, $\ell$) of the problem. This has previously been established explicitly for various CFTs. For a continuous group, the result is given by eq.~\!\eqref{eq:LieS}.}

    \item \textit{We reformulate the calculation of SREE, bypassing the computation of charged moments, using orbifold CFT. This approach makes it clear why symmetry resolution requires having a non-anomalous symmetry, allowing us to define the orbifolding process. See \S\ref{sec:orb}.}
    \item \textit{We derive the symmetry-resolved entanglement spectra for the reduced density matrix corresponding to a single interval in $1+1$D CFT with a finite symmetry group. We revisit a similar problem for CFT with $U(1)$ symmetry and obtain an improved version of symmetry-resolved entanglement spectra, compared to \cite{Goldstein2018}. See eqs.~\eqref{eeu1:main},~\eqref{eeu1:main2},~\eqref{eeu1:main3} and~\eqref{eq:nlambda} for the estimates.}
    \item \textit{We prove an optimal and rigorous lower and upper bound on the integrated entanglement spectra, leveraging the BCFT approach along with Tauberian theorems. To that end, we prove the theorem.~\!\ref{TBCFT2} and \ref{TBCFT} for the case of finite group symmetry. A similar analysis is performed for compact Lie groups, in particular $U(1)$, leading to theorem.~\!\ref{TBCFT3} and theorem.~\!\ref{TBCFTU1}. We contrast this result with that of finite group and point out the presence of extra polynomial suppression in the eigenavalue and exponential suppression by the Casimir of the irrep under consideration. Several numerical checks is performed for the EE spectra in the abelian case and a perfect agreement is found. } 
\end{itemize}

The paper is organized as follows. In Section \ref{sec:setup}, we review the BCFT setup for the entanglement entropy and we adapt this approach to compute the leading order behavior of the SREE with a given symmetry $G$, both for finite and compact Lie groups. We go into detail of SREE for finite groups in Section \ref{sec:SREEfinite}, where we can exploit the BCFT method to evaluate the charged moments of the
reduced density matrix, by computing the action of the defect-line generating the symmetry on the boundary states of our theory. In Section \ref{sec:orb}, we can completely bypass the computation of the charged moments by relating the SREE to a partition function in the closed string channel using the orbifold CFT description of our problem. The decomposition of the entanglement spectrum in each symmetry sector is addressed in Section \ref{sec:spectrum}, where we prove lower and upper bounds on the (integrated) resolved spectrum for finite and compact Lie groups.
We draw our conclusions in Section \ref{sec:conclusions} and we include an Appendix \ref{sec:twisted boundary} with some remarks about $\omega$-twisted boundary states.

\textbf{Note added:} Using the framework of Algebraic QFT (AQFT), one can derive eq.~\!\eqref{eq:MAINRESULT}. In fact, a similar equation has been written down in the context of the thermal density matrix in ref.~\!\cite{Magan:2021myk}, using the results from ref.~\!\cite{Casini:2019kex}. This amounts to proving \cite{Magan:2021myk} eq.~\!\eqref{density}. The ref.~\!\cite{Magan:2021myk} further explains the connection between symmetry resolved results pertaining to the thermal density matrix and the entanglement structure of the vacuum in QFT in the AQFT formalism, which is inherently Lorentzian in nature. However, in our paper we leverage the usual formalism of Euclidean CFT i.e we assume unitarity and modular invariance. It would be exciting to explore the connection between the framework of AQFT and the usual framework of Euclidean CFT in future.

\section{Entangling surface and Symmetry resolution}\label{sec:setup}

\subsection{Definition and conceptual framework}
In quantum field theory the reduced density matrix (RDM) of a subregion $A$ of the Cauchy slice in a state is formally obtained by tracing out the states, supported on complementary region $A^c$. 
\begin{equation}
\rho_A:= \mathrm{Tr}_{A^c} |\psi\rangle \langle \psi|\,.
\end{equation} 
 In any continuum field theory description, the formal reduced density matrix $\rho_A$ is ill-defined and affected by ultraviolet divergences due to the degrees of freedom at the entangling surface, i.e.\! the space codimension $1$ boundary that exists between $A$ and its complementary part. The UV divergence creates an inherent obstruction in writing down an isomorphism between the full Hilbert space $\mathcal{H}$ and the tensor product of Hilbert spaces living on $A$ and $A^c$ i.e.\! $\mathcal{H}_A\times \mathcal{H}_{A^c}$. With the absence of an isomorphism between $\mathcal{H}$ and $\mathcal{H}_A\times \mathcal{H}_{A^c}$, there is no natural way to trace out the complementary part. Therefore, the formal definition of $\rho_A$ falls apart. This implies that we need to specify what we are doing at the entangling surface. Mathematically, we introduce a family of maps $\imath$, labelled by some parameter $a$:
\begin{equation}
    \imath_a: \mathcal{H} \to \mathcal{H}_{A,a}\times \mathcal{H}_{A^c,a}\,.
\end{equation}
Given a state $|\psi\rangle\in\mathcal{H}$, we consider the state $\imath|\psi\rangle$. The reduced density matrix supported on the region $A$ is given by \cite{Ohmori2015} 
\begin{equation}
    \rho_{A,a}:=\mathrm{Tr}_{B,a} \left[\imath|\psi\rangle \langle \psi| \imath^\dagger\right]\,.
\end{equation}

The above discussion is not entirely for mathematical purists. The map $\imath_a$ has a physical implication, when we think of a lattice system, whose continuum limit yields the QFT under consideration. For example, in the Ising chain in $1+1$D, the endpoints of an interval $A$ can have arbitrary spin or can be restricted to have fixed spin. The existence of such a map is not obvious, for example the ref. \cite{Hellerman:2021fla} argued for non-existence of such map in the presence of gravitational anomaly. In what follows, we assume that our theory is free of such anomalies.

In the rest of the discussion, we specialize on $1+1$D conformal field theories on an infinite line and the case where $A$ is a path-connected single interval. The mapping $\imath_a$ which regularizes the UV divergence amounts to introducing small disks of radius $\epsilon$ around the entangling points \cite{Cardy2004}. If we insist on not breaking the conformal invariance, we need to impose conformal invariant boundary conditions at the endpoints, which we can label by $\alpha,\beta$. The corresponding spacetime geometry can be conformally mapped to that of an annulus of width $$W=2\log(\ell/\varepsilon)+O(\epsilon)\,,$$ We remark that we could also consider different geometries, like an interval in a finite size system, or at finite temperature: in all of these cases, the worldsheet for $\rho_A$ can still be mapped into an annulus by a conformal map, and the only change is in the width $W$, for more details we refer to \cite{Cardy2004}. 
In terms of the modular parameter $q=e^{2\pi i \tau}=e^{-2\pi^2/W}$, the RDM can be written as
\begin{equation}
\rho_A=\frac{q^{L_0-\frac{c}{24}}}{Z(q)},
\end{equation}
 where $Z(q):=\mathrm{Tr} q^{L_0-\frac{c}{24}}$ is the \textit{open string} partition function of a CFT with boundary conditions $\alpha,\beta$, and $L_0$ is the Virasoro zero mode. This trace is meant to be over all the open string states of the CFT under consideration. The trace of $n$th power of the RDM can be written as
 \begin{equation}\label{eq:open}
     \mathrm{Tr}\rho_A^n=\frac{Z(q^n)}{(Z(q))^n}, \quad Z(q):=\mathrm{Tr} q^{L_0-\frac{c}{24}}\,.
 \end{equation}
The R\'enyi and entanglement entropy are given by 
 \begin{equation}\label{eq:Renyi}
     S_n(q):=\frac{1}{1-n}\log \frac{Z(q^n)}{(Z(q))^n},\quad
     S_{vN}(q):=\lim_{n\to 1} S_n(q).
 \end{equation}
 By using the modular transformation (which is commonly known as \textit{open-closed duality}), we can rewrite the partition function in eq.~\eqref{eq:open} in terms of the dual modular parameter $\tilde{q}=e^{-2W}$ as
 \begin{equation}\label{eq:closed}
    Z(q^n)=\widetilde{\bra{a_1}}     \tilde{q}^{\frac{1}{n}(L_0-\frac{c}{24})}    \widetilde{\ket{a_2}}\,.
 \end{equation}
Here $\widetilde{\ket{a_1}}$ and $\widetilde{\ket{a_2}}$ are the Cardy boundary states. 

The main focus of our paper is to symmetry resolve the entanglement entropy for a CFT with global symmetry $G$. Thus the mapping $\imath$ needs to be compatible with the symmetry group $G$. In terms of Boundary Conformal Field Theory (BCFT), it requires choosing symmetry-preserving boundary conditions i.e. $G$ invariant Cardy boundary states. Furthermore, we will choose these boundary states to be simple (elementary); these boundary states are characterized by the existence of a unique ground state in the open string channel. For a more detailed discussion, see \cite{Choi:2023xjw}, where the notion of symmetry-preserving elementary boundary conditions is discussed comprehensively along with its extension to the case of non-invertible symmetry. In what follows, we will always mean \textit{elementary boundary states} by \textit{boundary states} unless otherwise mentioned. 

Before explaining what it means to have symmetry-preserving boundary conditions, let us explain formally how to compute the symmetry-resolved entanglement entropy. It is obtained from the following quantity, defined as
\begin{equation}
    \mathcal{Z}(q^n,r):=\mathrm{Tr}[\mathrm{\Pi}_r \rho_A^n],
\end{equation}
 as the $n$-th power of density matrix projected in the $r$-representation. Here $\Pi_r$ is the projection operator onto the $r$-representation. We can easily verify that $\mathcal{Z}(q,r)$ is a non-negative number satisfying  
 \begin{equation}
     \sum_r \mathcal{Z}(q,r)=1,
 \end{equation}
 i.e. it can be interpreted as the probability of finding $r$ as an outcome of the measurement of the charge. 
 In terms of the projected partition function, the SREE is given by 
 \begin{equation}\label{eq:defSREE}
     S_n(q,r):=\frac{1}{1-n}\log\frac{\mathcal{Z}(q^n,r)}{\mathcal{Z}^n(q,r)}, \qquad S_{vN}(q,r)=\lim_{n\to 1}S_n(q,r).
 \end{equation}
 We observe that in the replica limit $n\to 1$, the total entanglement entropy $S_{vN}(q)$ admits the following decomposition \cite{Lukin,nielsen2002quantum}
 \begin{equation}\label{eq:decomp}
   S_{vN}(q) =-\sum_r\mathcal{Z}(q,r)\log \mathcal{Z}(q,r)+\sum_r\mathcal{Z}(q,r)S_{vN}(q,r)=:S^{\rm num}+S^c\,.
 \end{equation}
 The first contribution, $S^{\rm num}$, is known as number entropy and it takes into account the entropy due to the fluctuations of the charge in the subsystem $A$, while the configurational entanglement entropy, $S^c$, is the weighted sum of the entropies in each irrep $r$.
 We further define the \textit{charged moment} of the RDM as
 \begin{equation}\label{eq:chargedM}
    Z(q^n,g):= \mathrm{Tr} \left[U^A(g)q^{n(L_0-\frac{c}{24})}\right]
 \end{equation}
 where $U^A(g)$ is the unitary matrix corresponding to the element $g \in G$, acting on states, supported on $A$ (a similar definition of the \textit{charged R\'enyi entropies} was introduced in \cite{Belin:2013uta}). This definition is inherently in the language of open string channel. The quantity $\mathcal{Z}(q^n,r)$ can be obtained from the charged moment by leveraging the character orthogonality theorem e.g.\! for a finite group $G$, we can explicitly write 
 \begin{equation}\label{eq:zcal}
     \mathcal{Z}(q^n,r)=\frac{d_r}{|G|}\sum_{g\in G}\chi_r^*(g)\frac{Z(q^n,g)}{(Z(q))^n},
 \end{equation}
 where $d_r$ is the dimension of the $r$-representation, $|G|$ is the order of the group and $\chi_r(g)$ is the group character. 

 The requirement of the symmetry-preserving boundary condition is hidden in the definition of charged moment in eq.~\eqref{eq:chargedM}. In the language of BCFT, the eq.~\eqref{eq:chargedM} computes the partition function in the open string channel with an insertion of a topological symmetry defect $U^A(g)$. We require the symmetry defect to end topologically on the boundary of the interval. This implies we need to choose a boundary condition which allows us to place the symmetry defect with endpoints being topological. In the closed string channel, the topological endability amounts to choosing $G$ invariant boundary states $\widetilde{\ket{a_1}}$ and $\widetilde{\ket{a_2}}$. For invertible symmetry (in this paper by symmetry, we always mean invertible symmetries corresponding to some group $G$), the topological endability is always equivalent to having a $G$ invariant boundary state. For noninvertible symmetry they are not equivalent \cite{Choi:2023xjw}, it is more natural to require topological endability condition to define symmetry resolution if possible. 

 In the absence of 't Hooft anomaly, there is no dynamical obstruction to have such $G$ invariant Cardy boundary state. In case, we have such a $G$ invariant Cardy boundary state, eq.~\eqref{eq:chargedM} is well defined and can be expressed in the closed string channel in the following way:
  \begin{equation}\label{eq:chargedMC}
    Z(q^n,g)=
    {}_g \widetilde{\bra{a_1}} \tilde{q}^{\frac{1}{n}(L_0-\frac{c}{24})}   \widetilde{\ket{a_2}} {}_g\,,\quad q=e^{-2\pi^2/W}\,,\tilde{q}=e^{-2W}
 \end{equation}
 where $\widetilde{\ket{a_1}}_g,\widetilde{\ket{a_2}}_g  $ are the boundary states in the defect Hilbert space twisted by $U^A(g)$. When $g$ is the Identity element, we write
 \begin{equation}
     \widetilde{\ket{a_i}} {}_e \equiv \widetilde{\ket{a_i}} \,,
 \end{equation}
$\widetilde{\ket{a_i}}$ are $G$ invariant Cardy boundary states.

 \subsection{A universal formula for charged moment \& SREE }\label{sec:universal}

\subsubsection*{Finite Group}
We consider $1+1$D CFT with central charge $c$, having a finite symmetry group $G$. As explained in the previous subsection, the insertion of the symmetry defect in the open string channel, from the perspective of closed string channel, is an amplitude between two twisted boundary states. The closed string states that propagate in this channel belong to the twisted Hilbert space. The ground state energy of this twisted Hilbert space is given by $\Delta^{(g)}_0-c/24$, which is strictly greater than that of ($-c/24$) the untwisted sector~\cite{Chang:2018iay, Pal:2020wwd}. This implies that in $q\to 1$ i.e. $\tilde{q}\to 0$ limit the charged moments behave as
     \begin{equation}\label{eq:leading}
       Z(q^n,g)\equiv   {}_g \widetilde{\bra{a_1}} \tilde{q}^{\frac{1}{n}(L_0-\frac{c}{24})}   \widetilde{\ket{a_2}} {}_g \sim \widetilde{\bra{a_1}} \Delta_*\rangle \langle\Delta_* \widetilde{\ket{a_2}}\tilde{q}^{\frac{1}{n}\left(\Delta_*-c/24\right)}\,,\quad \Delta_*\geqslant \Delta^{(g)}_0>0\,,\ \text{for}\ g\neq e\,.
     \end{equation}
The leading contribution to $Z(q^n,e)$ in $\tilde{q}\to 0$ (i.e. $q\to 1$) comes from the ground state of the untwisted sector, i.e. the state with $L_0$ eigenvalue $0$. Consequently, we have
 \begin{equation}\label{universalCM}
     \lim_{q\to 1}\frac{Z(q^n,g)}{Z(q^n,e)}=\delta_{g,e}\,.
 \end{equation}

 In what follows, we will omit the $e$, when we insert trivial symmetry defect and denote $Z(q^n,e)$ as $Z(q^n)$, $Z(q,e)$ as $Z(q)$.

 At this point, we use the asymptotic expression for the charged moments in eq.~\eqref{universalCM} to obtain
 \begin{equation}\label{eq:gen_finite}
  \mathcal{Z}(q^n,r)\underset{\eqref{eq:zcal}}{=} \frac{d_r}{|G|}\sum_{g\in G}\chi_r^*(g)\frac{Z(q^n,g)}{(Z(q))^n} \underset{\eqref{eq:leading}}{\sim }  \frac{d^2_r}{|G|}\frac{Z(q^n)}{(Z(q))^n}\,,\ q\to 1
 \end{equation}
 The subleading terms in the $q\to 1$ limit are exponentially suppressed. 
Finally, using eq.~\eqref{eq:Renyi} and eq.~\eqref{eq:defSREE}, we explicitly derive
 \begin{equation}\label{eq:finiteGres}
    \lim_{q\to 1} \left(S_n(q,r)-S_n(q)\right)\underset{\eqref{eq:gen_finite}}{=}\log\frac{d_r^2}{|G|}.
 \end{equation}
 In L.H.S., the first term is the symmetry-resolved R\'enyi entropies in the $r$-representation, the second term is the total R\'enyi entropy, which does not depend on the charge sector. The R.H.S. explicitly depends on the dimension of the $r$-representation and on the order of the group $|G|$. This is one of our main results. This result resembles the one found for the continuous group in Wess-Zumino-Witten models in the expression for symmetry-resolved R\'enyi entropies: the leading order term in the subsystem size $\ell$ (recall $q=e^{-\pi^2/\log(\ell/\varepsilon)})$, i.e. the total R\'enyi entropy, does satisfy the equipartition of entanglement while the $\mathcal{O}(\ell^0)$ depends
on the specific representation of the group, breaking the equipartition.
From the result for $\mathcal{Z}(q,r)$, one can also compute the number entropy defined in eq.~\eqref{eq:decomp} as
\begin{equation}
    S^{\mathrm{num}}=-\sum_r\mathcal{Z}(q,r)\log \mathcal{Z}(q,r)=\log |G|-\sum_r\frac{d^2_r}{|G|}\log d_r^2.
\end{equation}

The effect of the boundary condition on the entangling surface cancels out in the L.H.S. of \eqref{eq:finiteGres}. However, they do appear in the expression for symmetry-resolved R\'enyi entropies. In fact, as mentioned earlier, the definition of such symmetry-resolved quantity comes with a notion of boundary conditions. In the subsequent sections, we will give a careful analysis of the symmetry-resolved entanglement entropy with physical boundaries in some concrete examples.

 \subsubsection*{Compact Lie group}

For the case of compact Lie group, let us consider the $U(1)$ symmetry. The extension to arbitrary compact Lie group is straightforward; if we denote by $\mathfrak{g}$ the Lie algebra of $G$, we can use the Cartan generators $h^b$ ($b=1,\dots , {\rm rank}~\mathfrak{g}$), i.e. the generators of a maximal commuting subalgebra of $\mathfrak{g}$, so that $\sum_{b}\theta_{b} h^{b}$ is an element of the Cartan subalgebra $\mathfrak{h} \subset \mathfrak{g}$, instead of an arbitrary element of $\mathfrak{g}$, exploiting the fact that any element of $\mathfrak{g}$ is conjugated to an element of $\mathfrak{h}$. The reason behind this is that for any element $\sum_{a}z_{a}J^{a} \in \mathfrak{g} $, we can find $g \in G$ satisfying $\sum_{a}z_{a}J^{a} = g^{-1} \sum_{b}\theta_{b} h^{b} g$ for some $\sum_{b}\theta_{b} h^{b} \in \mathfrak{h}$. Now, for fields admitting a free field representation, the currents $j(x)$ belonging to Cartan subalgebra are dual to a
simple free boson operator $\partial \phi$. Thus effectively, the consideration of arbitrary compact Lie group boils down to a computation involving $U(1)$ symmetry \cite{Milekhin2021}. For simplicity, we consider a free compact boson with Luttinger parameter $K=1$ (see section \ref{app:spectrum} for more details about it).

The projected partition function for the charged sector $Q$ is given by (the irreps of $U(1)$ are naturally labeled by the charge $Q$, hence we switch notation here and denote the irrep label $r$ as $Q$)
 \begin{equation}\label{eq:gen_u1}
 \begin{split}
      \mathcal{Z}(q^n,Q)&= \frac{1}{(Z(q))^n}\int_{-1/2}^{1/2}d\theta\  e^{-2\pi i Q\theta} Z\left(q^n,e^{2\pi i\theta}\right).
 \end{split}
 \end{equation}
Using $q=e^{-\beta}$, with $\beta=\pi^2/\log(\ell/\varepsilon)$, in the $\beta\to 0$ limit the charged moments behave as (for simplicity, we consider $\widetilde{\ket{a_1}}=\widetilde{\ket{a_2}}=:\widetilde{\ket{a}}.$)
\begin{equation}\label{eq:charged_asymp}
    Z\left(q^n,e^{2\pi i\theta}\right) \sim |\widetilde{\bra{a}}\theta\rangle|^2 e^{\frac{4\pi^2}{n\beta}(c/24-\theta^2/2)}.
\end{equation}
Here $|\theta\rangle$ is the ground state of the twisted Hilbert space, where twisting is implemented by $g=e^{2\pi i\theta}$. The ground state energy is given by $\theta^2/2-c/24$, leading to the above expression.

Now let us define 
$$Z_Q(q^n):=\mathcal{Z}(q^n,Q)(Z(q))^n\,.$$ 

Plugging eq.~\eqref{eq:charged_asymp} in eq.~\eqref{eq:gen_u1}, we derive 
\begin{equation}
\begin{split}
   Z_Q(q^n)&\sim e^{\frac{4\pi^2}{n\beta}c/24} \int_{-1/2}^{1/2} d\theta\ e^{-2\pi i Q\theta} |\widetilde{\bra{a}}\theta\rangle|^2  e^{\frac{4\pi^2}{n\beta}(c/24-\theta^2/2)}\\
   &\sim |\widetilde{\bra{a}}0\rangle|^2 \int_{-\infty}^{\infty } d\theta\ e^{-2\pi i Q\theta} e^{\frac{4\pi^2}{n\beta}(c/24-\theta^2/2)} = e^{\frac{4\pi^2}{n\beta}c/24-\frac{n\beta Q^2}{2}} \sqrt{\frac{n\beta}{2\pi}} \,,\ 
\end{split}
\end{equation}
which leads to the following behavior in the $q\to 1 $ limit
\begin{equation}
    \mathcal{Z}(q^n,Q)\sim  |\widetilde{\bra{a}}0\rangle|^2 e^{-\frac{4\pi^2}{\beta}c/24(n-1/n)-\frac{n\beta Q^2}{2}} \sqrt{\frac{n\beta}{2\pi}}.
\end{equation}

Hence, using the definition of the SREE in eq.~\eqref{eq:defSREE}, we have ($q=e^{-\pi^2/\log(\ell/\varepsilon)}$)
 \begin{equation}\label{eq:u1SREE}
     S_n(q,Q)- S_n(q)\underset{\ell/\varepsilon\to\infty}{\to}\frac{\log (n)}{1-n}-\frac{1}{2} \log \left(\frac{2}{\pi} \log \left(\frac{\ell}{\varepsilon }\right)\right)+O(\ell^0).
 \end{equation}
This result shows that, for a CFT with $U(1)$ symmetry, the SREE does not depend on $Q$, i.e. it satisfies equipartition, in agreement with the findings of Ref. \cite{sierra, DiGiulio2022}.

Starting from the results for the $U(1)$ case, we can compute the projected partition function for any compact Lie group $G$. It reads
\begin{equation}\label{eq:LieZ}
    \mathcal{Z}(q^n,r)=
    d_r\int d\mu (g)\chi^*_r(g) \mathrm{Tr}[\rho_A^n e^{i\sum_{a}\theta_a h^{a}}]
    %\simeq d_r\int d\mu (g)\chi^*_r(g) \mathrm{Tr}[\rho_A^n e^{2\pi i\sum_{a}\alpha_aJ_{r,a}}],
\end{equation}
which is very similar to eq.~\eqref{eq:gen_finite}, where now the discrete sum over $g$ is replaced by the integral over the Haar measure of $G$.

 For a semi-simple Lie group $G$, with conjugation invariant  integrand $f$ i.e $f(kgk^{-1})=f(g)$, we have (for example, see appendix B of \cite{Henning:2017fpj})
\begin{equation}\label{eq:measure}
   \int\ d\mu (g)\ f(g)=\frac{1}{|W(G,T)|}\int\ \prod_{i=1}^\mathfrak{r}\ d\theta_i\ \prod_{\alpha\in rt_+(G)}\left(1-e^{2\pi i\alpha_i\theta_i}\right)\left(1-e^{-2\pi i\alpha_i\theta_i }\right)\ f(g)
   %\underset{\theta\to 0}{\sim} \prod_{i=1}^r\ d\theta_i\ \prod_{\alpha\in rt_+(G)} 4\pi^2 (\theta\cdot\alpha)^2
\end{equation}
Here $rt_+(G)$ refers to the set of positive roots and $|W(G,T)|$ is the order of the Weyl group. $\mathfrak{r}$ is the rank of the group. $T$ denotes the maximal torus with $\theta_i$ being the coordinates on it.

 We recall the general argument presented in the beginning of this section about how a calculation for the compact group can be reduced to a computation involving Cartan subgroup, which is nothing but collection of $U(1)$s. Now using the result for the $U(1)$, as given by eq.~\eqref{eq:charged_asymp} we find for level $1$, 
 \begin{equation}\label{chargedR}
     \mathrm{Tr}[\rho_A^n e^{i\sum_{a}\theta_a h^{a}}]\underset{\beta\to 0}{\sim} Z(q^n)e^{-\frac{2\pi^2 }{n\beta}F^{-1}_{ij}\theta_i \theta_j}\,
 \end{equation}
 Here $F^{-1}_{ij}=K(h^i,h^j)$ with $h^i$ being the Cartan generators in Chevalley basis and $K$ being the killing form defined by $K:=\frac{1}{2g}\mathrm{Tr}\left[\text{ad}X\text{ad}Y\right]$, where $g$ is dual coxeter number. If $(\lambda_1,\lambda_2,\cdots \lambda_{\mathfrak{r}})$ is the highest weight representation $\lambda$, then $h^i|\lambda\rangle=\lambda_i |\lambda\rangle$ with $\lambda_i\in\mathbb{Z}$. 

 \begin{remark}\label{remark1}
     The generalization to level $k$ is straightforward by changing $\beta^{-1} \to k\beta^{-1}$ in eq.~\!\eqref{chargedR}. Thus for the level $k$, we just need to do this replacement in what follows below.
 \end{remark}
 
We further recall the expression for the character in terms of highest weight vector $\lambda$, corresponding to the irrep $r$:
\begin{equation}\label{eq:char}
    \chi_{r}(g)= \frac{\sum_w \text{sgn}(w) e^{2\pi i w(\theta)_i\left(\lambda+\rho \right)_i}}{\prod_{\alpha\in rt_+(G)} {\left(e^{\pi i\alpha_i\theta_i}-e^{-\pi i\alpha_i\theta_i }\right)}}\,,
\end{equation}
where the sum in the numerator is over the Weyl group and $\rho=1/2\sum_{\alpha\in rt_+(G)}\alpha$ is the Weyl vector.

We plug in eq.~\!\eqref{eq:char} and the asymptotic expression for the charged moment in the eq.~\! \eqref{eq:LieZ} to derive 
\begin{equation}\label{eq:znq_r1}
\begin{split}
   &\mathcal{Z}(q^n,r)\sim   Z(q^n)\,|\widetilde{\bra{a}}0\rangle|^2 d_r\,\frac{1}{|W(G,T)|}\times\\
   &\times\sum_{w} \text{sgn}(w) \int \prod_{i=1}^\mathfrak{r}\ d\theta_i\ \prod_{\alpha\in rt_+(G)}\left(e^{\pi i\alpha_i \theta_i }-e^{-\pi i\alpha_i\theta_i}\right)\   e^{-\frac{2\pi^2 }{n\beta}F^{-1}_{ij}\theta_i\theta_j -2\pi i w(\theta)_i\left(\lambda+\rho \right)_i} \,.
\end{split}
\end{equation}
For a fixed $w$, we can do a change of variable from $\theta\to w(\theta)$ to realize the summand in the second line of \eqref{eq:znq_r1} is $w$ invariant. Hence we have
\begin{equation}\label{eq:znq_r}
\begin{split}
   &\mathcal{Z}(q^n,r)\sim   Z(q^n)\,|\widetilde{\bra{a}}0\rangle|^2 d_r\int \prod_{i=1}^\mathfrak{r}\ d\theta_i\ \prod_{\alpha\in rt_+(G)}\left(e^{\pi i\alpha_i\theta_i }-e^{-\pi i\alpha_i\theta_i }\right)\   e^{-\frac{2\pi^2 }{n\beta}F^{-1}_{ij}\theta_i\theta_j -2\pi i \theta_i\left(\lambda+\rho \right)_i} 
\end{split}
\end{equation}

Now we perform another change of variable $2\pi\theta_i= \sqrt{n\beta}\phi_i$:

\begin{equation}
\begin{split}
    &\frac{\mathcal{Z}(q^n,r)}{Z(q^n)}\sim |\widetilde{\bra{a}}0\rangle|^2d_r \left(\frac{n\beta}{4\pi^2}\right)^{\mathfrak{r}/2}\times \\
&\times\left[\int_{-\infty}^{\infty} \prod_{i=1}^\mathfrak{r}\ d\phi_i\ \prod_{\alpha\in rt_+(G)}\left(e^{\frac{i}{2}\sqrt{n\beta}\alpha_i\phi_i }-e^{-\frac{i}{2}\sqrt{n\beta}\alpha_i \phi_i }\right)\ e^{-\frac{1}{2}F^{-1}_{ij}\phi_i\phi_j -\imath\sqrt{n\beta} \phi_i(\lambda+\rho)_i}\right]
    \end{split}
\end{equation}
At this point, we take $\beta\to 0$ limit and evaluate the integral in saddle point approximation with the saddle given by $\phi_i=- iF_{ij}\sqrt{n\beta}(\lambda+\rho)_j$ and  obtain 

\begin{equation}
\begin{split}\label{eq:NEW}
    &\frac{\mathcal{Z}(q^n,r)}{Z(q^n)}\sim |\widetilde{\bra{a}}0\rangle|^2d_r e^{-\frac{n\beta}{2}||\lambda+\rho||^2}\left(\frac{n\beta}{2\pi}\right)^{\mathfrak{r}/2}\left(\frac{n\beta}{2\pi}\right)^{| rt_+(G)|} \frac{\prod_{\alpha\in rt_+(G)} \alpha\cdot (\lambda+\rho)}{\prod_{\alpha\in rt_+(G)} \alpha\cdot \rho} \times \\
    &\times\underbrace{\left(\prod_{\alpha\in rt_+(G)} 2\pi \alpha\cdot \rho\right)\sqrt{F}e^{-\frac{n\beta}{2}||\rho||^2}}_{\text{irrep independent and non-singular in}\ \beta}
    \end{split}
\end{equation}
Here $|rt_+(G)|$ is the number of positive roots and the inner product $(\ \cdot\ )$ and the norm $||\cdot||$ are taken using the quadratic form $F_{ij}$.

Recalling the basic results about dimension $d_r$ and quadratic casimir $C_2(r)$ of the irrep $r$ along with the formula for the  dimension $d_G$ of the Lie algebra corresponding to a semi-simple compact Lie group:
\begin{equation}
    d_r=\frac{\prod_{\alpha\in rt_+(G)} \alpha\cdot (\lambda+\rho)}{\prod_{\alpha\in rt_+(G)} \alpha\cdot \rho}\,,\quad d_G=\mathfrak{r}+2|rt_+(G)|\,,\quad C_2(r)=\left(||\lambda+\rho||^2-||\rho||^2\right)
\end{equation}
we write eq.~\!\eqref{eq:NEW} in the following form and perform the integral:

    \begin{equation}
\begin{split}\label{answerRoot}
     &\frac{\mathcal{Z}(q^n,r)}{Z(q^n)}\sim 
%|\widetilde{\bra{a}}0\rangle|^2\, d_r^2\,\ e^{-n\beta C_2(r)}\ \left(\frac{n\beta}{2\pi}\right)^{d_G/2} \times \\
    % &\quad\quad\quad\times\left(\prod_{\alpha\in rt_+(G)} \frac{\alpha\cdot \rho}{2\pi}\right)e^{-\frac{n\beta}{2}||\rho||^2}\left[\left(\frac{1}{\sqrt{2\pi}}\right)^{\mathfrak{r}}\int_{-\infty}^{\infty} \prod_{i=1}^\mathfrak{r}\ d\phi_i\ e^{-\frac{1}{2}||\phi -\imath\sqrt{n\beta} (\lambda+\rho)||^2}\right] \\
|\widetilde{\bra{a}}0\rangle|^2\,d_r^2\, e^{-\frac{n\beta}{2}C_2(r)}\left(\frac{n\beta}{2\pi}\right)^{d_G/2}\underbrace{\left(\prod_{\alpha\in rt_+(G)} 2\pi \alpha\cdot \rho\right)\sqrt{F}e^{-\frac{n\beta}{2}||\rho||^2}}_{\text{irrep independent and non-singular in}\ \beta}
    \end{split}
\end{equation}
Here $F:=\text{det}(F_{ij})$.Therefore, we can write the analogous of eq.~\eqref{eq:u1SREE} for a generic compact Lie group $G$, by substituting $\beta=\frac{\pi^2}{\log(\ell/\varepsilon)}$:
\begin{equation}\label{eq:LieS}
\begin{split}
     & S_n(q,r)- S_n(q)\underset{\ell/\varepsilon\to\infty}{=}\frac{d_G\log (n)}{2(1-n)}-\frac{d_G}{2} \log \left(\frac{2}{\pi} \log \left(\frac{\ell}{\varepsilon }\right)\right)+2\log d_r+O(\ell^0),
\end{split}
 \end{equation}
where we highlight the term $2 \log(d_r)$,  which explicitly depends
on the specific irrep of $G$, breaking equipartition at $O(\ell^0)$. This result is in agreement with the analysis done for WZW models in \cite{calabrese2021symmetryresolved}. 

\begin{remark}
   We note that, following the remark \ref{remark1}, the level dependence in \eqref{eq:LieS} will come out as an additive term $-d_G/2\log k$, consistent with \cite{calabrese2021symmetryresolved}.
\end{remark}

A point worth remarking in context of comparing our result to \cite{calabrese2021symmetryresolved}, no other saddle contributes in our computation other than the one we consider. This can be traced back to the fact that we always work in a basis such that the Dynkin labels of the highest weight vector are integers. 
Thus we expect that in the case of $SU(N)$, with the choice of killing form as appeared in \cite{calabrese2021symmetryresolved}, we must have 
\begin{equation}
\frac{\sqrt{N}(2\pi)^{\frac{N^2+N}{2}-1}}{\prod_{k=1}^{N-1}k!}
\underset{\text{\cite{calabrese2021symmetryresolved}}}{=}
vol(SU(N)) \underset{?}{=} (2\pi)^{d_G}\left(\sqrt{F}\prod_{\alpha\in rt_+(G)} 2\pi \alpha\cdot \rho\right)^{-1}
\end{equation}
Indeed the above can be verified by noting that for $SU(N)$ we have  $d_G=N^2-1$, $|rt_+(G)|=N(N-1)/2$ and $F^{-1}$ is a tridiagonal $(N-1)\times (N-1)$ matrix with $F^{-1}_{11}=2$,$F^{-1}_{12}=-1$ and $F^{-1}_{21}=-1$ and so on (see appendix 13.A of \cite{DiFrancesco1997}, $F^{-1}$ is given by the Cartan matrix $A$ there) such that the $F=\text{det}(F)=\frac{1}{N}$. The simple roots are given by $\alpha_i= F^{-1}_{ij}\omega_j$ where $\omega_j$ are fundamental roots. Finally $\rho=\sum_i\omega_i$ i.e $\rho=(1,1,\cdots 1)$ in the fundamental root basis. We further have $\prod_{\alpha\in rt_+(G)}\alpha\cdot\rho= \prod_{k=1}^{N-1}k!$, the most convenient way to derive this (see section $13.3.2$ of the book \cite{DiFrancesco1997}) is to use the orthonormal basis in $N$ dimension, where the positive roots are given by $e_i-e_j$ with $i<j$ and $\rho$ is given by $\rho=\sum_i\omega_i=\sum_i \rho_i e_i$ with $\rho_i=\frac{N-1}{2}-i+1$. Thus we have $\prod_{\alpha\in rt_+(G)}\alpha\cdot\rho= \prod_{1\leqslant i<j\leqslant N}(j-i)=G(n+1)=\prod_{k=1}^{N-1}k!$, where $G(n)$ is the Barnes function.

\section{Symmetry resolution of the entanglement for finite groups}\label{sec:SREEfinite}

\subsection{Symmetry generated by Verlinde line}\label{subsec:dH}

Global symmetries can be defined in terms of topological defects.
Here, we focus on a particular class of topological defect, called a Verlinde line, which can be systematically constructed in a similar way to a Cardy boundary.
If a Verlinde line is constructed from a simple current, this Verlinde line generates a symmetry.\footnote{Otherwise, the Verlinde line generates a non-invertible symmetry, which is defined in terms of the fusion category.}
Thus, we call this line the symmetry defect.

A simple current $J$ is a primary field, whose fusion with any other primaries leads to only one primary, $J\times\mu=\lambda$.
The alternative definition is given by the existence of a conjugate $J\times J^c = \bb{I}$.
There are only a finite number of primaries in rational CFT (RCFT), i.e.
there exists an integer $n$ such that
\begin{equation}
J^n \times \lambda = \lambda.
\end{equation}
Therefore, a set of simple currents generates an abelian group,
\begin{equation}
\bb{Z}_{n_1} \times \bb{Z}_{n_2} \times \cdots.
\end{equation}
One characterization of the simple current is given in terms of the quantum dimension as
\begin{equation}
d_J = 1.
\end{equation}
Note that in WZW models based on simple lie algebra except for $E_8$ at level $2$,
all simple currents are related to outer automorphisms \cite{DiFrancesco1997}. 
\footnote{
The simple currents in minimal models can be understood from the coset construction.
}

A symmetry defect that we are interested in can be described as the Verlinde line associated with a simple current,
\begin{equation}
I_\lambda = \sum_\mu \fr{S_{\lambda \mu}}{S_{0 \mu}} ||\mu||,
\end{equation}
where $||\mu||$ is defined by 
\begin{equation}
||\mu|| := \sum_{N,\bar{N}} \ket{\mu;N} \otimes \ket{\bar{\mu};\bar{N}}   \bra{\mu;N} \otimes \bra{ \bar{\mu};\bar{N} },
\end{equation}
where the state $\ket{\mu; N}$ is in the Verma module $\ca{V}_\mu$ and its level is labeled by $N$.
One can show using the fusion rule of simple currents that the symmetry defect forms a group.
A symmetry generated by the symmetry defect is
\begin{equation}\label{eq:action}
\lambda: v \to \ex{2\pi i Q_\lambda(\mu)} v, \ \ \ \ v \in \ca{V}_\mu \otimes \overline{\ca{V}}_{\mu},
\end{equation}
where the charge $Q_g(i) $ is defined by
\begin{equation}\label{eq:charge}
Q_g(i) \equiv h_g + h_i - h_{g(i)}  \mod 1.
\end{equation}
Or equivalently, this definition just comes from
\footnote{
Formally, the charged moments can be defined as a two-point function in the orbifold CFT,
\begin{equation}
Z_n(\alpha) = \braket{\ca{T}_{n,\alpha}(0) \bar{\ca{T}}_{n,\alpha}(l) },
\end{equation}
where $\ca{T}_{n,\alpha}$ is a composite of the twist operator $\sigma_n$ and the vertex operator $\ca{V}_{\alpha}$, that is, $\ca{T}_{n,\alpha} = \sigma_n  \cdot \ca{V}_{\alpha}$.
Here, we realize the monodromy (\ref{eq:monodromy}) by acting the vertex operator $\ca{V}_\alpha$ on the entangling surface. While this is the most common and arguably a useful way to compute charged moments, this approach suffers from several conceptual issues. As a starter, in the twist field approach, we cannot keep track of the lattice size dependence. The regulator is implicitly included in the normalization of the twist fields \cite{Lunin2001} and the dependence on the regulator is captured only  the leading order. It is also unclear how to see the effect of the boundary condition on the entangling surface in this twist operator approach. The approach in \cite{Jia:2023ais, Jia:2023dpt} might shed some light in this direction.
}
\begin{equation}\label{eq:monodromy}
\lambda(\ex{2\pi i}z)\times \mu(0) = \ex{-2\pi i Q_\lambda(\mu)} \lambda(z) \times \mu(0),
\end{equation}
where $\lambda(z)$ is the simple current and $\mu(z)\in \ca{V}_\mu \otimes\overline{\ca{V}}_{\mu}$.
One important property of the charge may be $S_{g(i),j} = \ex{2\pi i Q_g(j)} S_{ij}$ \cite{Schellekens1990},
which leads to eq.~\eqref{eq:action}.
Some other properties can be found in \cite{Brunner2004}.

The corresponding defect Hilbert space has a simple form,
\begin{equation}
\ca{H}_\lambda = \bigoplus_{\mu,\nu} N_{\lambda \mu}^\nu \ca{V}_\mu \otimes \overline{\ca{V}_\nu},
\end{equation}
which can be shown by using the Verlinde formula. The constant $N_{\lambda \mu}^\nu$ is the fusion matrix.
The symmetry defect associated with the global symmetry generated by the simple current $J$ is simply given by the Verlinde line $I_J$.
By the fusion rule of the simple current, the defect Hilbert space is
\footnote{
Another derivation can be found in \cite{Brunner2004}.
}
\begin{equation}\label{eq:hil}
\ca{H}_J = \bigoplus_{\mu} \ca{V}_{\mu} \otimes \overline{\ca{V}_{J \mu}}.
\end{equation}

%%%%%%%%%%%%%%%%%%%%%%%%%%%%%%%%%%%%%%%%%%%%%%%%%%%%%%%%%%%%%%%%%%%%%%%%%%%%%%%%%%%%%%%%%%%%%%
\subsection{Symmetry-resolved entanglement entropy}
%%%%%%%%%%%%%%%%%%%%%%%%%%%%%%%%%%%%%%%%%%%%%%%%%%%%%%%%%%%%%%%%%%%%%%%%%%%%%%%%%%%%%%%%%%%%%%

Let $G$ be a group generated by a simple current and $\widetilde{\ket{b}}$ be a boundary state invariant under $g \in G$.
The charged moments are given by
\begin{equation}
Z(q^n, g) = {}_g \widetilde{\bra{b}} \ti{q}^{\fr{1}{n} (L_0+\bar{L}_0 -\fr{c}{12})} \widetilde{\ket{b}}_g.
\end{equation}
The conditions for the boundary state are given by
\begin{equation}\label{eq:conds}
\pa{L_n -  \bar{L}_{-n}}\widetilde{\ket{b}}=0, \ \ \ \ 
\pa{W_n - \pa{-1}^h \ti{W}_{-n}} \widetilde{\ket{b}}=0,
\end{equation}
where $W_n$ are the modes of the current $W(z)$ associated with the chiral algebra $\ca{A}$ and $\ti{W}_{-n}$ are those of the antiholomorphic chiral algebra $\ti{\ca{A}}$ while $h$ is the conformal dimension of $W_n$. The left equation derives from the gluing condition $T=\bar{T}$ along the boundary, where $T$ ($\bar{T}$) is the (anti)-holomorphic component of the stress-energy tensor. The right equation is due to a similar constraint on the current, $W=(-1)^h\bar{W}$. 
Moreover, the boundary states should be invariant under $g \in G$ to be compatible with the open-string duality.

In diagonal RCFT, one can construct some solutions to these equations by the Cardy construction,
\begin{equation}
\widetilde{\ket{\lambda}} = \sum_\mu \fr{S_{\lambda, \mu}}{ \sqrt{S_{0\mu}}} \kett{\mu},
\end{equation}
where $S_{\lambda \mu}$ is the modular $S$ matrix  and $\kett{\mu}$ is the Ishibashi state.
The sum runs over primaries associated with the chiral algebra.
One can twist this boundary state by replacing the modular $S$ matrix with the generalized modular $S$ matrix,
\begin{equation}\label{eq:twS}
\widetilde{\ket{\lambda}}_J = \sum_\mu \fr{S_{\lambda, \mu}(J)}{ \sqrt{S_{0\mu}}} \kett{\mu}_J,
\end{equation}
where the sum is taken over scalar primaries in the twisted Hilbert space $\ca{H}_J$,
and the generalized modular $S$ matrix $S_{\lambda, \mu}(J)$ is the modular $S$ matrix for a one-point torus block with the primary field $J$ (see \cite{Bantay1998}).
We call the state $\kett{\mu}_J$ the twisted Ishibasi state, which is the Ishibashi state included in the twisted Hilbert space and should be distinguished from $\kett{\mu}$.

We would like to give a comment on how to construct a symmetric boundary state.
It is not well-understood how to construct a symmetric boundary in a systematic way.
Moreover, it is still non-trivial if there really exists a symmetric boundary state in a given CFT even if the symmetry is free of `t Hooft anomaly.\footnote{
A symmetric boundary state is possible only if the symmetry is free of `t Hooft anomaly (see \cite{Choi2023} and wherein).
Can we always find a symmetric boundary state if the symmetry is non-anomalous?
The answer is still unknown but it might be true from investigations of concrete examples \cite{Wang2023,Han2017,Smith2021,Smith2020,Tong2022,Li2022,Zeng2022,Wang2022}.}
Therefore, our approach is to search for a symmetric boundary from all Cardy states, which are boundary states that we can systematically construct.
There are more boundary states that can be systematically constructed, called the $\omega$-twisted boundary state (see App. \ref{sec:twisted boundary}).
For example, the Neumann boundary state in a free boson CFT is an $\omega$-twisted boundary state, where the boundary state is twisted by the charge conjugation along the boundary.
The Neumann boundary state is invariant under the $u(1)$ symmetry, therefore,  we can use this $\omega$-twisted boundary as the boundary condition on the entangling surface.

\subsubsection{Examples}
\paragraph{$\mathbb{Z}_2$ in Ising model:}
In the Ising model, there are three Virasoro primary operators/fields,
the identity operator $\mathbb{I}$ with $h=\bar h=0$,
the energy density field $\epsilon $ with $h=\bar h=1/2$ and
the spin field $\sigma $ with $h=\bar h=1/16$. The nontrivial fusion rules among these primary operators are given by 
\begin{equation}
    \epsilon \times\epsilon \sim \mathbb{I}\,,\ \epsilon \times \sigma \sim \sigma \,,\ \sigma\times \sigma \sim \mathbb{I}+\epsilon\,.
\end{equation}
From the fusion rule, we can read that $\epsilon$ is a simple current. The $\epsilon$ generates the usual spin flip $\mathbb{Z}_2$ symmetry under which $\mathbb{I}$ and $\epsilon $ are even while $\sigma$ is odd. We can verify this explicitly by evaluating the charges of the operators using eq.~\eqref{eq:charge}:
\begin{equation}
Q_\epsilon(\mathbb{I}) = 0, \ \ \ \ 
Q_\epsilon(\epsilon) =0, \ \ \ \
Q_\epsilon(\sigma) = \frac{1}{2}.
\end{equation}

The Ising characters are given by
\begin{equation}\label{eq:chi_ising}
    \begin{split}
   \chi_{0}(q)=\frac{1}{2}\left(\sqrt{\frac{\theta_3(q)}{\eta(q)}}+\sqrt{\frac{\theta_4(q)}{\eta(q)}}\right), \, \chi_{1/2}(q)=\frac{1}{2}\left(\sqrt{\frac{\theta_3(q)}{\eta(q)}}-\sqrt{\frac{\theta_4(q)}{\eta(q)}}\right),\,  \chi_{1/16}(q)=\sqrt{\frac{\theta_2(q)}{2\eta(q)}},  
    \end{split}
    \end{equation}
where the $\theta_i$ are the Jacobi theta functions, and $\eta$ is the Dedekind eta function. Under modular transformation, they are transformed according to the modular-S transformation, which is given by  matrix (in the ordered basis $\{0,\epsilon,\sigma  \}$) 
\begin{equation}\label{eq:S}
S=
  \left(
    \begin{array}{ccc}
     \frac{1}{2}  & \frac{1}{2} & \frac{1}{\sqrt{2}} \\
     \frac{1}{2}  & \frac{1}{2} & -\frac{1}{\sqrt{2}} \\
      \frac{1}{\sqrt{2}} & -\frac{1}{\sqrt{2}} & 0 \\
    \end{array}
  \right).
  \end{equation}

  Explicitly, we have
  \begin{equation}
      \chi_{s}(q^n)= \sum_t S_{st} \chi_t\left(\tilde{q}^{\frac{1}{n}}\right)\,;\quad \ s,t\in \{0,1/2,1/16\}.
  \end{equation}
The Cardy boundary states are given by 
\begin{equation}\label{eq:boundarysigma}
    \begin{split}
        \widetilde{\ket{\mathbb{I}}}&=\frac{1}{\sqrt{2}}\ket{\mathbb{I}}\!\rangle+\frac{1}{\sqrt{2}}\ket{\epsilon}\!\rangle+\frac{1}{2^{1/4}}\ket{\sigma}\!\rangle\,, \\
        \widetilde{\ket{\epsilon}}&=\frac{1}{\sqrt{2}}\ket{\mathbb{I}}\!\rangle+\frac{1}{\sqrt{2}}\ket{\epsilon}\!\rangle-\frac{1}{2^{1/4}}\ket{\sigma}\!\rangle\,, \\
          \widetilde{\ket{\sigma}}&=\ket{\mathbb{I}}\!\rangle-\ket{\epsilon}\!\rangle\,. 
    \end{split}
\end{equation}
Clearly, $\widetilde{\ket{\sigma}}$ is symmetric under $\mathbb{Z}_2$ and it is the only symmetric one.

The charged moment corresponding to the insertion of trivial symmetry ($e$) defect is given by
    \begin{equation}\label{eq:triv}
Z(q^n,e):= \widetilde{\bra{\sigma}}\tilde{q}^{\frac{1}{n}(L_0-c/24)} \widetilde{\ket{\sigma}}=\chi_{0}(\tilde{q}^{\frac{1}{n}})+\chi_{1/2}(\tilde{q}^{\frac{1}{n}})\,.
\end{equation}
Here we express the charged moment using the modular transformed frame i.e. the closed string channel, hence it is an explicit function of $\tilde{q}$. Note the definition comes with an explicit choice of boundary states in the closed string channel. The second equality follows from using the expression for $\widetilde{\ket{\sigma}}$ in terms of Ishibashi states.

Now we would like to insert the nontrivial symmetry defect and evaluate $Z(q^n,g)$, where $g$ is the nontrivial element of the group $\mathbb{Z}_2$. This amounts to twisting the $\widetilde{\ket{\sigma}}$ by introducing the nontrivial $\mathbb{Z}_2$ symmetry defect in the closed string channel and defining a twisted Cardy state. This twisted boundary state is a linear combination of twisted scalar Ishibashi states coming from the $\mathbb{Z}_2$ twisted sector of the Ising model. We recall that the field content of $\mathbb{Z}_2$ twisted sector of Ising model is given by $(h,\bar h)=$ $(0,1/2)$, $(1/2,0)$ and $(1/16,1/16)$. The twisted Ishibashi coming from the operator with $h=\bar h=1/16$  is the only scalar operator; let us denote it as $\ket{\sigma}\!\rangle_\epsilon$. Thus the twisted Cardy state must be given by
\begin{equation}
\widetilde{\ket{\sigma}}_\epsilon=\alpha \ket{\sigma}\!\rangle_\epsilon \,,
\end{equation}
for some constant $\alpha\in\mathbb{C}$. The corresponding charged moment with insertion of nontrivial $\mathbb{Z}_2$ symmetry defect is given by 
\begin{equation}
Z(q^n,g)= \alpha^2\chi_{1/16}(\tilde{q}^{\frac{1}{n}})=\frac{\alpha^2}{\sqrt{2}} \chi_{0}(q^n)  -\frac{\alpha^2}{\sqrt{2}} \chi_{1/2 }(q^n)\,,
\end{equation}
where the second equality follows from using $S$-modular matrix as given in eq.~\eqref{eq:S}. Since the rightmost expression is in the open string channel, we can read off the action of symmetry operator on the $h=0$ and $h=1/2$ operator from there as 
\begin{equation}
    U(g)|h=0\rangle = \frac{\alpha^2}{\sqrt{2}} |h=0\rangle \,,\ U(g)|h=1/2\rangle = -\frac{\alpha^2}{\sqrt{2}} |h=1/2\rangle
\end{equation}
where $g$ is the nontrivial $\mathbb{Z}_2$ element. Requiring $\pm \frac{\alpha^2}{\sqrt{2}} \in\{-1,1\}$ and $\alpha\in\mathbb{R}$, we obtain $\alpha=2^{1/4}$ and we have 
\begin{equation}\label{eq:non}
Z(q^n,g)= \sqrt{2}\chi_{1/16}(\tilde{q}^{\frac{1}{n}})= \chi_{0}(q^n) - \chi_{1/2 }(q^n).
\end{equation}
Now using eqs.~\eqref{eq:non} and~\eqref{eq:triv}, specifically the closed string channel expressions, we have
\begin{equation}\label{eq:supp}
    \frac{Z(q^n,g)}{Z(q^n,e)}= \frac{\sqrt{2}\chi_{1/16}(\tilde{q}^{\frac{1}{n}})}{\chi_{0}(\tilde{q}^{\frac{1}{n}})+\chi_{1/2}(\tilde{q}^{\frac{1}{n}})} \underset{\tilde{q}\to 0}{\to} 0\,.
\end{equation}
Thus the projected partition function is given by
\begin{equation}
\begin{split}
\mathcal{Z}(q^n,\text{even})&= \frac{1}{2}\left(\frac{Z(q^n,e)}{(Z(q))^n}+\frac{Z(q^n,g)}{(Z(q))^n}\right)\underset{\tilde{q}\to 0}{\sim}  \frac{1}{2}\frac{Z(q^n,e)}{(Z(q))^n}   \,,\\
\mathcal{Z}(q^n,\text{odd})&= \frac{1}{2}\left(\frac{Z(q^n,e)}{(Z(q))^n}-\frac{Z(q^n,g)}{(Z(q))^n}\right)\underset{\tilde{q}\to 0}{\sim}  \frac{1}{2}\frac{Z(q^n,e)}{(Z(q))^n} \,.
\end{split}
\end{equation}
Here the limit follows from using~\eqref{eq:supp}. One can also deduce the above directly from the open string channel (again using eqs.~\eqref{eq:non} and~\eqref{eq:triv}, but now the explicit expression in the open string channel) by noting that 
\begin{equation}\label{eq:res_ising}
    \mathcal{Z}(q^n,\text{even})= \frac{\chi_0(q^n)}{(Z(q))^n}\,,\ \mathcal{Z}(q^n,\text{odd})= \frac{\chi_{1/2}(q^n)}{(Z(q))^n}\,.
\end{equation}

\paragraph{$\mathbb{Z}_k$ in $u(1)_k$:}
We consider WZW model with current algebra given by $u(1)_k$ chiral algebra. The $u(1)_k$ chiral algebra has $2k$ irreducible representations and we label them as elements from the set
\begin{equation}
\ca{I} = \{  (\lambda)| \lambda=0,1,\cdots,2k-1 \}\,.
\end{equation}
The modular $S$ matrix has the following form,
\begin{equation}
S_{\lambda \mu} = \fr{1}{\sqrt{2k}} \ex{-\fr{\pi i}{k} \lambda \mu}.
\end{equation}

The fusion rule is given by 
\begin{equation}
    (\lambda_1) \times (\lambda_2) = (\lambda_1+\lambda_2)
\end{equation}
where $\lambda_1+\lambda_2$ should be thought of as an integer modulo $2k$. Clearly, using eq.~\eqref{eq:charge}, we can see that  the $\bb{Z}_k$ symmetry is constructed from the simple currents $\{(\lambda_{inv})|\lambda_{inv}=0,2,4,\cdots, 2k-2\}$. We label the symmetry defects as $I_{(\lambda_{inv})}$.

The $I_{\lambda_{inv}}$-invariant boundary state can be found in the $\omega_c$-twisted Cardy states where $\omega_c$ is the charge conjugation (see App. \ref{sec:twisted boundary}),
\begin{equation}
\widetilde{\ket{\pm}} = \fr{(2k)^{\fr{1}{4}}}{\sqrt{2}} \pa{ \kett{0;\omega_c} \pm \kett{k;\omega_c}  }.
\end{equation}
Note that this boundary state is known as the Neumann boundary state.

The possible Ishibashi states in the twisted Hilbert space $\ca{H}_\lambda$ are $\kett{\fr{\lambda}{2} ;\omega_c}_\lambda$ and  $\kett{\fr{\lambda}{2}+k;\omega_c}_\lambda$.
\begin{equation}\label{eq:twistedu1}
\widetilde{\ket{\pm}}_{\lambda_{inv}} = \fr{(2k)^{\fr{1}{4}}}{\sqrt{2}} \bigg( \kett{\lambda_{inv}/2  ;\omega_c}_{\lambda_{inv}} \pm \kett{\lambda_{inv}/2+k ; \omega_c}_{\lambda_{inv}}  \bigg).
\end{equation}
We determine the coefficient $\fr{(2k)^{\fr{1}{4}}}{\sqrt{2}}$ by requiring that $|c_\lambda|$ is $1$, where $c_\lambda$ appears in the expression for 
$Z(q^{n},I_{\lambda_{inv}})=\sum_\lambda c_\lambda \chi^{(ex)}_\lambda(q^n)$. This is analogous to how we determined a similar coefficient for the Ising model.  

The charged moments are given by 
\begin{equation}
Z(q^n,I_{\lambda_{inv}})={}_{\lambda_{inv}} \widetilde{\bra{\pm}}\tilde{q}^{\frac{1}{n}(L_0-c/24)}\widetilde{\ket{\pm}}_{\lambda_{inv}} = \sum_{\mu \in \text{even}} \ex{-\fr{2\pi i\mu \lambda_{inv}}{2k} }  \chi_\mu^{(ex)} \pa{q^n}.
\end{equation}
Again the definition of charged moments comes with a choice of boundary state when there is trivial defect and then we need to twist it accordingly to obtain a result for $\lambda_{inv}\neq 0$. One can see from the closed string channel expression that $Z(q^n,I_{\lambda_{inv}} )$ for $\lambda_{inv} \neq 0$ are exponentially suppressed in the $\tilde{q}\to 0$ i.e. $\ell/\varepsilon \to \infty$ limit. This leads to 
\begin{equation}\label{u1kr}
  \ca{Z}(q^n,s) \underset{\tilde{q}\to 0}{\sim }\frac{1}{2}\frac{Z(q^n,I_0)}{(Z(q))^n} \,,\quad  s=\{\text{even},\text{odd}\}\,.
  \end{equation}

The projected partition is given by 
\begin{equation}\label{eq:Zu1}
\mathcal{Z}(q^n,s) =\fr{1}{k} \sum_\lambda \ex{\fr{2\pi i}{2k} s\lambda} \frac{Z(q^n,I_\lambda)}{(Z(q))^{n}} = \frac{\chi_s^{(ex)}\pa{q^n}}{(Z(q))^{n}},  \ \ \ \ \ \ s=\{0,2,4,... \}.
\end{equation}
One can verify eq.~\eqref{u1kr} from the above expression directly.

\paragraph{$\mathbb{Z}_2$ in $su(2)_{2k}$:}

Let us consider the $\bb{Z}_2$ symmetry in $su(2)_{2k}$ Wess-Zumino-Witten models, in which there are $2k+1$ irreducible representations labelled by elements from the following set
\begin{equation}
\ca{I} = \{ (\lambda)| \lambda = 0, 1/2,1, 3/2, \cdots, k    \},
\end{equation}
whose conformal dimension is given by $h_\lambda = \fr{\lambda(\lambda+1)}{2k+2}$.
The modular $S$ matrix has the following form,
\begin{equation}\label{eq:SSU}
S_{\lambda \mu} = \sqrt{\fr{2}{2k+2}} \sin \fr{   \pi(2\lambda+1)(2\mu+1)  }{2k+2    }.
\end{equation}

The fusion rule reads
\begin{equation}
    (\lambda_1) \times (\lambda_2) = \bigoplus_{\lambda=|\lambda_1-\lambda_2|; \lambda+\lambda_1+\lambda_2\in\mathbb{Z}}^{\text{Min}\left(\lambda_1+\lambda_2,2k-\lambda_1-\lambda_2\right)} \ (\lambda).
\end{equation}
In particular, for all $(\lambda)$, we have 
\begin{equation}
    (\lambda) \times (k) = \left(k-\lambda \right)\,.
\end{equation}
The above implies that $(k)$ is a simple current and generates the  $\bb{Z}_2$ symmetry. Using eq.~\eqref{eq:charge}, we can find
\begin{equation}
    Q_{(\lambda)}= \lambda\ \text{mod}\ 1\,.
\end{equation}
Thus $(\lambda)$ with $\lambda\in\mathbb{Z}$ are even states and
the $\bb{Z}_2$-invariant Cardy state is $\widetilde{\ket{k/2}}$.
The charged moments with insertion of trivial defect (we denote it as $I_0$ since it is generated by the trivial simple current $(0)$)  are given by
\begin{equation}
Z\pa{q^n,I_0 }:=\widetilde{\bra{k/2}} \tilde{q}^{\frac{1}{n}(L_0-\frac{c}{24})}\widetilde{\ket{k/2}}=\sum_{\mu = 0,1,\cdots,k} \chi_\mu\pa{\tilde{q}^{\frac{1}{n}}},
\end{equation}
Here again the definition of $Z\pa{q^n,I_0 }$ comes with the boundary condition $\widetilde{\ket{k/2}}$; this is manifest in the expression for the closed string channel. The final equality follows from using eq.~\eqref{eq:SSU} and we obtain an expression in the open string channel. We remark that $\chi$ is not the Virasoro character here, rather it is the character corresponding to $su(2)_k$ algebra.

Now we would like to evaluate the charged moment with insertion of non-trivial $\bb{Z}_2$ defect. This amounts to figuring out the twisted Cardy state. The twisted Cardy state is given by 
\begin{equation}
\widetilde{\ket{k/2}}_{(k)} = (k+1)^\fr{1}{4} \ket{k/2}\!\rangle_{(k)}.
\end{equation}
The coefficient $(k+1)^\fr{1}{4}$ can be determined by requiring that $|c_n|$ is $1$, where $c_n$ appears in the expression for 
$Z(q^{n},I_k)=\sum_\lambda c_\lambda \chi_\lambda(q^n)$. This is analogous to how we determined a similar coefficient for the Ising model. 

The charged moment with insertion of non-trivial $\bb{Z}_2$ defect ($I_k$) is given by
\begin{equation}
Z\pa{q^n,I_k } ={}_{(k)}\widetilde{\bra{k/2}} \tilde{q}^{\frac{1}{n}(L_0-\frac{c}{24})}\widetilde{\ket{k/2}}_{(k)}= \sqrt{k+1}\chi_{k/2}(\tilde{q}^\frac{1}{n})=\sum_{\mu = 0,1,\cdots,k} \ex{\pi i \mu} \chi_\mu\pa{\tilde{q}^{\frac{1}{n}}}.
\end{equation}

From the closed string channel expression, one can easily see that  $Z\pa{q^n,I_k } $ is exponentially suppressed in the $\tilde{q}\to 0$ i.e $\ell/\varepsilon \to \infty$ limit. This leads to 
\begin{equation}
  \ca{Z}(q^n,s) \underset{\tilde{q}\to 0}{\sim }\frac{1}{2}\frac{Z(q^n,I_0)}{(Z(q))^n} \,,\quad  s=\{\text{even},\text{odd}\}\,.
  \end{equation}
One can also verify this by computing the projected partition function
\begin{equation}
\begin{aligned}
\ca{Z}(q^n,s) = \sum_\mu  \frac{\chi_\mu\pa{q^n}}{(Z(q))^n}\,, \ \ \ \ \ \ \ \ \ \ 
&\left\{
    \begin{array}{ll}
    \mu \in \text{even}   ,& \text{if } s=\text{even}   ,\\
    \mu \in \text{odd}  ,& \text{if }  s=\text{odd} .\\
    \end{array}
  \right.\\
\end{aligned}
\end{equation}

%%%%%%%%%%%%%%%%%%%%%%%%%%%%%%%%%%%%%%%%%%%%%%%%%%%%%%%%%%%%%%%%%%%%%%%%%%%%%%%%%%%%%%%%%%%%%%
\paragraph{Remarks about anomalous symmetry:} 
%%%%%%%%%%%%%%%%%%%%%%%%%%%%%%%%%%%%%%%%%%%%%%%%%%%%%%%%%%%%%%%%%%%%%%%%%%%%%%%%%%%%%%%%%%%%%%

Here, we claim that there is no boundary state in defect Hilbert space if the symmetry is anomalous.
If a given symmetry is anomalous, 
all states in the corresponding defect Hilbert space obey a spin selection rule.
For example, in $\bb{Z}_2$ case, the spin selection rule is given by \cite{Lin2019}
\begin{equation}
\begin{aligned}
s&=\left\{
    \begin{array}{ll}
    \fr{\bb{Z}}{2}   ,& \text{if non-anomalous $\bb{Z}_2$}   ,\\
    \fr{1}{4}+ \fr{\bb{Z}}{2}  ,& \text{if anomalous $\bb{Z}_2$}   .\\
    \end{array}
  \right.\\
\end{aligned}
\end{equation}
Therefore, one can immediately find that the defect Hilbert space does not contain the vacuum primary,
which implies that the charged moments are always suppressed.
We can also claim that since there are no scalar primaries in the defect Hilbert space,
the possible Ishibashi state does not exist.
This means that there is no boundary state in the defect Hilbert space if the symmetry is anomalous.
For example, in $su(2)_1$ WZW model, there are two irreducible representations,
\begin{equation}
\ca{I} = \{ (\lambda)| \lambda = 0, \fr{1}{2}  \}.
\end{equation}
The anomalous $\bb{Z}_2$ symmetry is generated by a simple current $(\fr{1}{2})$.
The defect Hilbert space is given by
\begin{equation}
Z= \chi_0(\tau) \chi_{\fr{1}{2}}(\bar{\tau}) + \chi_{\fr{1}{2}}(\tau) \chi_0(\bar{\tau}),
\end{equation}
which implies that there is no possible Ishibashi state in the twisted sector.
This is related to obstruction of gauging due to `t Hooft anomaly (see section \ref{sec:orb}).

There is another way to see this.
According to \cite{Choi2023},
{\it elementary} symmetric boundary states are possible only if the symmetry is free of `t Hooft anomaly.
In fact, this does not necessarily imply that there is no boundary state in the defect Hilbert space
because one can always find {\it non-elementary} symmetric boundary states (e.g., spontaneously fixed boundaries \cite{Fukusumi2021, Prembabu2022}) even if the symmetry is anomalous.
Thus, one may wonder if the anomalous symmetry defect can end on the spontaneously fixed boundary
(see the discussion about the implication from the strongly symmetric condition to the weakly symmetric condition in \cite{Choi2023}).
This is true in the sense that the anomalous symmetry defect can {\it locally} end on the spontaneously fixed boundary.
But globally, this is impossible.
The anomalous symmetry defect acts as a boundary-changing defect, therefore, we need another symmetry defect to close the boundary.
Alternatively, the spontaneously fixed boundary state does not live in the anomalous defect Hilbert space\footnote{
We would like to thank Brandon Rayhaun and Shu-Heng Shao for an important discussion about this issue.
}.
As a result, we can conclude that the symmetry-resolution cannot be applied to a system with anomalous symmetry (at least if we use the regularization procedure introduced in Section \ref{sec:setup}).
It would be interesting to comment that a similar conclusion can be obtained in a lattice model.
On a lattice, one cannot write a symmetry operator as a product of local unitaries if the symmetry is anomalous.\footnote{We would like to thank Shu-Heng Shao and Masataka Watanabe for telling us this argument.}
It implies that one cannot even define the (anomalous) symmetry resolved entanglement entropy on a lattice.
%We will discuss this issue in our future paper.

%%%%%%%%%%%%%%%%%%%%%%%%%%%%%%%%%%%%%%%%%%%%%%%%%%%%%%%%%%%%%%%%%%%%%%%%%%%%%%%%%%%%%%%%%%%%%%
%%%%%%%%%%%%%%%%%%%%%%%%%%%%%%%%%%%%%%%%%%%%%%%%%%%%%%%%%%%%%%%%%%%%%%%%%%%%%%%%%%%%%%%%%%%%%%
\section{Symmetry Resolution by Orbifold}\label{sec:orb}
%%%%%%%%%%%%%%%%%%%%%%%%%%%%%%%%%%%%%%%%%%%%%%%%%%%%%%%%%%%%%%%%%%%%%%%%%%%%%%%%%%%%%%%%%%%%%%
%%%%%%%%%%%%%%%%%%%%%%%%%%%%%%%%%%%%%%%%%%%%%%%%%%%%%%%%%%%%%%%%%%%%%%%%%%%%%%%%%%%%%%%%%%%%%%

The SREE is obtained from a weighted sum of open string partition functions with symmetry defect inserted, where weighting is done by the characters as given in eq.~\eqref{eq:zcal}.
Such a quantity also appears in an orbifold CFT. In fact, the partition function of an orbifold CFT can be thought of as the sum of partition functions of all twisted sectors in the original theory (see, for example, section $2$, especially $2.5$ of the ref. \cite{Collier2021} for a quick exposition of these concepts). 
In the path integral formalism, the partition function of a twisted sector is realized by inserting the symmetry defect. This observation hints at the possibility of finding a connection between SREE and the partition function of an orbifold CFT.
In this section, we will make this connection precise leading to a prescription for computing SREE. The advantage of this procedure is to bypass the evaluation of the charged Renyi entropy.

The key role is played by the boundary states in the orbifold CFT. We remind that given a CFT invariant under a discrete group $G$, it can be gauged to produce another CFT, known as $G$-orbifold.
In general, the boundary states in this orbifold CFT are given by the following (they are called \textit{fractional branes} in BCFT literature),
\begin{equation}\label{eq:frac}
\widetilde{\ket{b,r_N}} = \sum_{h \in \mathrm{N}} \fr{\chi^N_{r_\mathrm{N}}(h)}{\abs{\mathrm{N}}} \sum_{g \in C_G(h)} \fr{\sqrt{\abs{G}}}{\abs{C_G(h)}} g   \widetilde{\ket{b}}_h,
\end{equation}
where the boundary state $\widetilde{\ket{b}}$ %$\widetilde{\ket{b,r_N}}
%$
is invariant under the subgroup $\mathrm{N} \subset G$, $\widetilde{\ket{b}}_h$ is the state obtained via twisting by $h$ (note, $\widetilde{\ket{b}}_e\equiv \widetilde{\ket{b}} $ ) and $\chi^N_{r_N}(h)$ is the character for the irreducible representation $r_N$ of N.
$C_G(h)$ is the center of $h$ in $G$.
Note that if $\widetilde{\ket{b}}$ is the Cardy state of the parent CFT, the fractional brane is the Cardy state of the orbifold CFT. We emphasize that $\widetilde{\ket{b}}_h$ is state obtained in the original CFT by taking the boundary state $\widetilde{\ket{b}}$ and twisting by $h\in N$ while $\widetilde{\ket{b,r_N}}$ is a boundary state in the orbifold CFT and labeled by the irrep $r_N$ of the group $N$. 

Let us consider a particular class of boundary states, which satisfy
\begin{equation}
g\widetilde{\ket{b}} = \widetilde{\ket{b}}, \ \ \ \ \text{for any } g \in G.
\end{equation}
For this symmetric boundary state, $\mathrm{N}$ is equal to $G$.
Therefore, the fractional brane has a simple form,
\begin{equation}
\widetilde{\ket{b,r}} = \sum_{g \in G} \fr{\chi^G_{r}(g)}{\sqrt{\abs{G}}}  \widetilde{\ket{b}}_g.
\end{equation}
Using these fractional branes, we have
\begin{equation}\label{eq:orbapp}
\widetilde{\bra{b,r}}  \ti{q}^{\fr{1}{n} (L_0+\bar{L}_0 -\fr{c}{12})}  \widetilde{\ket{b,1}}
=
\fr{1}{\abs{G}} \sum_g \chi^{G*}_{r}(g) Z(q^n, g).
\end{equation}
In particular, if we restrict ourselves to an abelian group, this can be written as
\begin{equation}\label{eq:amplitude}
\widetilde{\bra{b,1}}   \check{g}_{r}    \ti{q}^{\fr{1}{n} (L_0+\bar{L}_0 -\fr{c}{12})}    \widetilde{\ket{b,1}}
=
\fr{1}{\abs{G}} \sum_g \chi^{G*}_{r}(g) Z(q^n, g),
\end{equation}
where $\check{g}_{r} \in  \check{G} (\cong G)$ is a quantum symmetry, which is a symmetry of the orbifold CFT that acts on the $g$-twisted sector states $\ket{i}_g$ as
\begin{equation}
\check{g}_{r}: \ket{i}_g \to \chi^{G*}_{r}(g)\ket{i}_g.
\end{equation}
That is, the symmetry-resolved entanglement entropy corresponds to an annulus partition function of the orbifold CFT.
The point is that unlike the symmetry resolution of a thermal state \cite{Pal2020a},
we have a choice of entangling surfaces.
One can choose the symmetrized boundary as the entangling surface, which implements the projection on the $G$-invariant states.
Thus, the sum over twisted sectors in the right hand side of Eq.~\eqref{eq:amplitude} obviously leads to the orbifold theory.

This connection gives a new approach to evaluate the symmetry-resolved entanglement entropy.
The process is the following:
\begin{enumerate}

\item Find a boundary state in the orbifold CFT, $\widetilde{\ket{b,1}}$, such that $g\widetilde{\ket{b}} = \widetilde{\ket{b}}$ for any $g \in G$.

\item Evaluate the closed string amplitude with the quantum symmetry group element $\check{g}_{r}$.

\end{enumerate}
Note that it is obvious that this procedure does work only if the symmetry is free of `t Hooft anomaly, which is an obstruction to gauging.
One advantage may be that it can keep us from evaluating the charged moments and allows us to directly evaluate the symmetry-resolved entanglement entropy.
This result represents a generalization to \textit{any} finite group and CFT of what has been done for the $\mathbb{Z}_2$ symmetry of the compactified boson in \cite{DiGiulio2022}. 
\footnote{The approach in \cite{DiGiulio2022} to avoid the computation of the charged moments used the relation between the charge and scaling dimension of the boundary primaries, the ones propagating in the open string channel. This relation is manifestly encoded in the WZW characters with fugacity turned on and hence can be found for $u(1)$ theories and other WZW models. In particular, in Eq.$42$ and below in \cite{DiGiulio2022}, it is noted that $\chi_q=q^{h_Q}\eta^{-1}$ and $h_Q$ is given as a function of $Q$. However in general, $h$ and the label of the irrep (which is $Q$ for $U(1)$) is not obviously related and finding such relation is a challenging task for an arbitrary finite group and CFT. For instance, already in the Ising CFT it is not clear to us how this identification can be done.}
%In their  approach, we need the knowledge of the charge of the boundary primaries, which is non-trivial except for $u(1)$ theory and other WZW models. One can immediately see this non-triviality in the Ising CFT. 

%%%%%%%%%%%%%%%%%%%%%%%%%%%%%%%%%%%%%%%%%%%%%%%%%%%%%%%%%%%%%%%%%%%%%%%%%%%%%%%%%%%%%%%%%%%%%%
\subsection{$\bb{Z}_2$ in Ising model}
%%%%%%%%%%%%%%%%%%%%%%%%%%%%%%%%%%%%%%%%%%%%%%%%%%%%%%%%%%%%%%%%%%%%%%%%%%%%%%%%%%%%%%%%%%%%%%
Let us consider the Ising model as an example.
There are three Cardy states in the $\bb{Z}_2$ gauged theory,

\begin{equation}\label{eq:gaugedIsing}
\begin{aligned}
(+):& \ \ \ \widetilde{\ket{0}}_D = \fr{1}{\sqrt{2}}\ket{0}\!\rangle - \fr{1}{\sqrt{2}}\ket{\epsilon}\!\rangle  + \fr{1}{2^{\fr{1}{4}}}\ket{\sigma}\!\rangle_\epsilon, \\
(-):&\ \ \ \widetilde{\ket{\epsilon}}_D = \fr{1}{\sqrt{2}}\ket{0}\!\rangle - \fr{1}{\sqrt{2}}\ket{\epsilon}\!\rangle  - \fr{1}{2^{\fr{1}{4}}}\ket{\sigma}\!\rangle_\epsilon, \\ 
(f):& \ \ \ \widetilde{\ket{\sigma}}_D = \ket{0}\!\rangle + \ket{\epsilon}\!\rangle. \\
\end{aligned}
\end{equation}
The first two are constructed from the invariant Cardy state $\widetilde{\ket{\sigma}}$ by eq.~\eqref{eq:frac}.  By $\ket{\sigma}\!\rangle_\epsilon$ we mean twisting the Ishibasi state $\ket{\sigma}\!\rangle$ by nontrivial element of $\mathbb{Z}_2$ i.e $\ket{\sigma}\!\rangle_{g}\equiv \ket{\sigma}\!\rangle_\epsilon$ when $g$ is nontrivial, see the paragrapph following \eqref{eq:twS}.

Explicitly, we have two representation of $\mathbb{Z}_2$, the trivial one, denoted by $r=+$ (this is denoted as $r=1$ in the general discussion) and nontrivial one, denoted by $r=-$. In the notation of general discussion from earlier part of this section, we have
\begin{equation}
\widetilde{\ket{\sigma}}\equiv \widetilde{\ket{b}}\,,\quad 
\widetilde{\ket{0}}_D \equiv \widetilde{\ket{b,+}}\equiv \widetilde{\ket{b,1}}\,,\quad 
\widetilde{\ket{\epsilon}}_D \equiv \widetilde{\ket{b,-}} 
\end{equation}

If $\check{g}_{-}$ is the nontrivial element of the $\check{G}=\mathbb{Z}_2$, we have 
\begin{equation}
    \check{g}_- \widetilde{\ket{0}}_D =\widetilde{\ket{\epsilon}}_D 
\end{equation}

One can further see that the closed string amplitudes are given by 
\begin{equation}
\widetilde{\bra{b,1}}  \ti{q}^{\fr{1}{n} (L_0+\bar{L}_0 -\fr{c}{12})}  \widetilde{\ket{b,1}}\equiv {}_D \widetilde{\bra{0}} \ti{q}^{\fr{1}{n}(L_0 + \bar{L}_0-\fr{c}{12})}  \widetilde{\ket{0}}_D
= \chi_0\pa{q^n  }\,.
\end{equation}
and 
\begin{equation}
\widetilde{\bra{b,-}}  \ti{q}^{\fr{1}{n} (L_0+\bar{L}_0 -\fr{c}{12})}  \widetilde{\ket{b,1}}\equiv {}_D \widetilde{\bra{\epsilon}} \ti{q}^{\fr{1}{n}(L_0 + \bar{L}_0-\fr{c}{12})}  \widetilde{\ket{0}}_D
= \chi_{1/2}\pa{q^n  }\,.
\end{equation}

% \begin{equation}
% {}_D \widetilde{\bra{s}} \ti{q}^{\fr{1}{n}(L_0 + \bar{L}_0-\fr{c}{12})}  \widetilde{\ket{0}}_D
% =
% {}_D \widetilde{\bra{0}} \check{g}_q \ti{q}^{\fr{1}{n}(L_0 + \bar{L}_0-\fr{c}{12})} \widetilde{\ket{0}}_D
% =
%  \chi_s\pa{q^n  }, \ \ \ \ \ s \in \{0,1/2 \},
% \end{equation}
% Here we have used the notation ${}_D\widetilde{\bra{1/2}}\equiv {}_D\widetilde{\bra{\epsilon}}$.

The above closed string amplitudes in the $\mathbb{Z}_2$ gauged theory reproduce the SREE as evident from the agreement of the amplitudes with the eq.~\eqref{eq:res_ising}. This verifies our approach. 
One can generalize this result to any CFT with non-anomalous $\bb{Z}_2$ symmetry by using the mapping of boundary states under $\bb{Z}_2$ gauging \cite{Fukusumi2021a, Ebisu2021}.

%%%%%%%%%%%%%%%%%%%%%%%%%%%%%%%%%%%%%%%%%%%%%%%%%%%%%%%%%%%%%%%%%%%%%%%%%%%%%%%%%%%%%%%%%%%%%%
\subsection{General Argument}
%%%%%%%%%%%%%%%%%%%%%%%%%%%%%%%%%%%%%%%%%%%%%%%%%%%%%%%%%%%%%%%%%%%%%%%%%%%%%%%%%%%%%%%%%%%%%%

One can simplify our result in eq.~\eqref{eq:orbapp} by taking $\varepsilon \to 0$ limit.
In this limit, only the vacuum state propagates in the closed string amplitude.
It is naturally expected that the vacuum in the orbifold CFT is unique if the vacuum of the parent CFT is unique.
In fact, one can show the uniqueness of the vacuum in the following way.
If the vacuum degenerates, the vacuum should be non-trivially transformed under the dual symmetry.
It implies that one can expand the orbifold vacuum by states in the twisted sectors of the parent CFT.
This is not possible because the lowest energy in the defect Hilbert space is strictly larger than the vacuum energy \cite{Chang2019}.
Therefore, we can argue
\begin{quote}
{\it
the vacuum of the orbifold CFT is unique if the vacuum of the parent CFT is unique.
}
\end{quote}
Using this fact and eq.~\eqref{eq:orbapp},
one can give the symmetry-resolved entanglement entropy in the $\varepsilon \to 0$ limit as
\begin{equation}
S_n(q,r) = \fr{c}{6}\pa{1+\fr{1}{n}} \log \fr{\abs{A}}{\varepsilon} + \log \fr{d_r^2}{\abs{G}} + \log \abs{\braket{\widetilde{a}|0}}^2.
\end{equation}
Here $|A|$ is the linear size of the subsystem.

\subsection{Symmetry resolution of the entanglement for continuous groups}
The goal of this subsection is to extend the formalism of previous subsections to continuous groups. The result is a generalization of the content of \cite{Goldstein2018}, which focused on the resolution of the $U(1)$ symmetry of the compactified boson. Our formalism applies to \textit{any} conformal field theory which admits a free field representation. As we have already stressed, the problem boils down to the evaluation of a closed string amplitude given a proper boundary state in the orbifold CFT. 
If we start from a state which is symmetric under the action of a continuous group $G$, $\widetilde{\ket{b}}$, we can construct the following boundary state in orbifold CFT
\begin{equation}
    \widetilde{\ket{b,r}}=d_r\int d\mu (g)\chi_r(g) e^{i\sum_{a}\alpha_aJ_{r,a}}\widetilde{\ket{b}}\,,
\end{equation}
where $d\mu (g)$ is the Haar measure of $G$
and $e^{i\sum_{a}\alpha_aJ_{r,a}}$ is an element of $G$ via the exponential map.
Using again the free field representation described in Section \ref{sec:universal}, we can explicitly write down the projected partition function as 
\begin{equation}
    \mathcal{Z}(q^n,r)=
    \widetilde{\bra{b,r}} \tilde{q}^{\frac{1}{n}(L_0-\frac{c}{24})}   \widetilde{\ket{b,1}}=
    d_r\int d\mu (g)\chi^*_r(g) \mathrm{Tr}[\rho_A^n e^{i\sum_{a}\alpha_aJ_{r,a}}].
\end{equation}
The right-hand side of the equation above agrees with the result found for WZW models in \cite{calabrese2021symmetryresolved} and resembles the result in eq. \eqref{eq:LieZ}. We stress that this result proves there exists a connection between the projected partition function $\mathcal{Z}(q^n,r)$ and the closed string amplitude $\widetilde{\bra{b,r}} \tilde{q}^{\frac{1}{n}(L_0-\frac{c}{24})}   \widetilde{\ket{b,1}}$ also for continuous groups. This is the main result of this section, which would allow us to compute the SREE bypassing the knowledge of the charged moments used in section \ref{sec:universal}.

\section{Entanglement spectrum}\label{sec:spectrum}
The entanglement entropy is a coarse-grained quantity, derived from the RDM $\rho_A$. However, $\rho_A$ encodes much more information. A relevant quantity characterizing the detailed structure of $\rho_A$ is the distribution of its eigenvalues $\lambda_i$, known as entanglement spectrum and defined as
\begin{equation}
P(\lambda):=\sum_i\delta(\lambda-\lambda_i).
\end{equation}
Here the sum is over the eigenvalues including the multiplicity. 

The symmetry-resolved entanglement spectrum is defined by refining the eigenvalue distribution, restricting the sum to a given irreducible representation of the symmetry group. In particular, we recall that the RDM $\rho_A$ has a block-diagonal decomposition:
\begin{equation}
    \rho_A =\bigoplus_{r}\rho_{A,r}
 \end{equation}
 where the sum is over the irreducible representation (labeled by $r$) of the group. 
 
Given $\rho_{A,r}$, there are two standard ways to define symmetry-resolved entanglement spectra, depending on whether we choose to normalize $\rho_{A,r}$ or not. Let us denote the normalized one as $\rho_{A,r}^{(N)}$:
\begin{equation}
    \rho_{A,r}^{(N)}:=\frac{\rho_{A,r}}{\mathrm{Tr}\ \rho_{A,r}} \,.
\end{equation}
We denote the eigenvalues of $\rho_{A,r}$ as $\lambda_{i,r}$ while the eigenvalues of $\rho_{A,r}^{(N)}$ are denoted as $\lambda^{(N)}_{i,r}$. Similarly we define $P_r$, the symmetry-resolved entanglement spectra coming from $\rho_{A,r}$ and $P^{(N)}_r$, the symmetry-resolved entanglement spectra coming from $\rho^{(N)}_{A,r}$.
\begin{equation}
P_r(\lambda):=\sum_i\delta(\lambda-\lambda_{i,r})\,, \quad P^{(N)}_r(\lambda):=\sum_i\delta(\lambda-\lambda^{(N)}_{i,r})\,.
\end{equation}
Again $r$ labels the irreducible representation (irrep) of the group. 

The goal of this section is to derive a formula for the entanglement spectrum of $\rho_{A,r}$ and $\rho^{(N)}_{A,r}$. This problem has been studied previously for a global $U(1)$ symmetry (where $r$ is in fact the charge $Q$) in \cite{Goldstein2018}, using the approach introduced by Calabrese and Lefevre in \cite{Lefevre_2008} to compute $P_r(\lambda)$. In particular, we derive
\begin{enumerate}
    \item a formula for finite group and a more explicit formula for compact Lie group including $U(1)$. For $U(1)$, the formula is an \textit{improvement} over that of \cite{Goldstein2018}. See eqs.~\eqref{eeu1:main},~\eqref{eeu1:main2},~\eqref{eeu1:main3} and~\eqref{eq:nlambda}.
    \item  the rigorous upper and lower bound on integrated $P_r$ and $P_{r}^{(N)}$. The technical results pertaining to BCFT are contained in theorem.~\!\ref{TBCFT} for the finite group, the remarks that follow the theorem, the generalization to compact Lie group as described in the  theorem.~\!\ref{TBCFTU1} and the remarks following that. The implications of these theorems in terms of entanglement spectra are described in the theorem.~\!\ref{TBCFT2} for finite group and the extension to compact Lie group in theorem.~\!\ref{TBCFT3} along with the remarks that follow.
\end{enumerate}

\subsection{Symmetry-resolved Entanglement spectra}
The BCFT spectra is related to eigenvalues $\lambda_i$ of $\rho_A$ via the following equation \cite{Alba:2017bgn}:
\begin{equation}\label{spectrumBCFT1}
    \lambda(\Delta)=\lambda_0 e^{-\beta(\Delta-c/24)}\,,\quad \beta\equiv \frac{\pi^2}{\log(\ell/\varepsilon)}\,,
\end{equation}
where $\lambda_0$ is the maximum eigenvalue of $\rho_A$. One can further derive $\lambda_0^{-1}=Z(\beta)$.
Now we restrict the above sum within irrep $r$ to obtain 
\begin{equation}\label{spectrumBCFT2}
    \lambda_r(\Delta_r)=\lambda_0 e^{-\beta(\Delta_r-c/24)}\,,\quad \beta\equiv \frac{\pi^2}{\log(\ell/\varepsilon)}\,.
\end{equation}
Here $\Delta_r$ denotes the scaling dimension of an operator transforming under the irrep $r$ of the group $G$, and $\lambda_r$ denotes the eigenvalue of the block $\rho_{A,r}$.

Furthermore, we can divide by the projected partition function i.e. $\mathrm{Tr}(\rho_{A,r})\equiv Z_r(\beta)$,  to obtain 
\begin{equation}\label{eq:normEig}
\lambda^{(N)}_r(\Delta)=\frac{\lambda_0}{Z_r(\beta)} e^{-\beta(\Delta_r-c/24)}\,,\quad \beta\equiv \frac{\pi^2}{\log(\ell/\varepsilon)}\,.
\end{equation}
Thus we identify $\frac{\lambda_0}{Z_r(\beta)}$ as the maximum eigenvalue of $\rho_{A,r}^{(N)}$ and denote it as $\lambda^{(N)}_{0,r}$. The eq.~\eqref{eq:normEig} becomes
\begin{equation}
\lambda^{(N)}_r(\Delta_r)=\lambda^{(N)}_{0,r} e^{-\beta(\Delta_r-c/24)}\,,\quad \beta\equiv \frac{\pi^2}{\log(\ell/\varepsilon)}\,.
\end{equation}
We note that $\beta(\Delta_r-c/24)=\log(\lambda_0/\lambda)=\log(\lambda^{(N)}_{0,r}/\lambda^{(N)})$. Hence we have 
\begin{equation}
    \int d\lambda'\ P_r(\lambda') =\int d\lambda'\ P^{(N)}_r(\lambda') = \int d\Delta'\ \rho_r(\Delta')
\end{equation}
where we need to impose the integration limit appropriately, depending on the variable of the integration. Now we derive an expression for $\rho_r(\Delta')$, which is the Inverse Laplace transformation of the projected partition function $Z_r(\beta)$. Since we will be interested in $\Delta\to\infty$ limit of density of states, this is captured by $Z_r(\beta)$ in the $\beta\to 0$ limit. This is the usual intuitive argument behind Cardy-like statements in CFT literature. \\

$\bullet$ For the finite group, we already derived $Z_r(\beta) \sim \frac{d_r^2}{|G|} Z(\beta)$.
Thus we deduce that
\begin{equation}
    \rho_r(\Delta)\sim \frac{d_r^2}{|G|}\rho(\Delta)\,.
\end{equation}

Therefore, for finite group, $P_r(\lambda)=d_r^2/|G|  P_{\text{full}}(\lambda)$, and $P^{(N)}_r(\lambda)=d_r^2/|G|  P_{\text{full}}(\lambda)$ where $P_{\text{full}}(\lambda)$ is the full entanglement spectra, as written down in \cite{Alba:2017bgn}.\\

$\bullet$ For the continuous compact Lie group of rank $p$, we have $p$ Cartan generators, and we can turn on fugacity for each of these Cartan directions and finally do a Haar integral to obtain a projected partition function. Let us explain this using the simplest setup, where the group is  $U(1)$. The generalization to compact Lie group of arbitrary rank $p$ group is straightforward by restricting to the maximal torus. \\

For a $U(1)$ compact boson with Luttinger parameter $K=1$, the projected partition function for $\beta\in \mathbb{R}_+$ is given by (the generalization to other values of $K$ is straightforward, introducing a factor of $K^{-1/2}$ in the final expression) 
\begin{equation*}
    Z_Q(\beta)=e^{\frac{4\pi^2}{\beta}c/24} \int_{-1/2}^{1/2} d\theta e^{-2\pi i Q\theta} e^{-\frac{4\pi^2}{\beta}\theta^2/2} |\widetilde{\bra{a}}\theta\rangle|^2.
\end{equation*}
In the $\beta\to 0$, assuming the $|\widetilde{\bra{a}}\theta\rangle|^2$ is a slowly varying function, we can do a saddle point analysis and consequently extend the integral from $-\infty$ to $\infty$. We obtain
\begin{equation}
\begin{split}
   Z_Q(\beta)&=e^{\frac{4\pi^2}{\beta}c/24} \int_{-1/2}^{1/2} d\theta\ e^{-2\pi i Q\theta} |\widetilde{\bra{a}}\theta\rangle|^2  e^{\frac{4\pi^2}{\beta}(c/24-\theta^2/2)}\\
   &\sim |\widetilde{\bra{a}}0\rangle|^2 \int_{-\infty}^{\infty } d\theta\ e^{-2\pi i Q\theta} e^{\frac{4\pi^2}{\beta}(c/24-\theta^2/2)} = |\widetilde{\bra{a}}0\rangle|^2e^{\frac{4\pi^2}{\beta}c/24-\frac{\beta Q^2}{2}} \sqrt{\frac{\beta}{2\pi}}. 
\end{split}
\end{equation}
 This is precisely the Eq.~\!$7$ of \cite{Goldstein2018} upon setting $K=1,n=1$ and $\beta=\frac{\pi^2}{\log(\ell/\varepsilon)}$ except the term $|\widetilde{\bra{a}}0\rangle|^2 $, whose presence cannot be captured by the use of twist operators as done in \cite{Goldstein2018}.  Now we do the inverse Laplace transformation to obtain $\rho_Q(\Delta)$:
\begin{equation}
\begin{split}
     \rho_Q(\Delta)=&|\widetilde{\bra{a}}0\rangle|^2 \int d\beta\  \sqrt{\frac{\beta}{2\pi}}e^{\frac{4\pi^2}{\beta}c/24-\frac{\beta Q^2}{2}+\beta(\Delta-c/24)} \\
&=|\widetilde{\bra{a}}0\rangle|^2 \Theta\left(\Delta_Q\right)\left[\frac{\sqrt{c} \sinh \left(2 \pi  \sqrt{\frac{c \Delta_Q }{6}}\right)}{2 \sqrt{3}  \Delta_Q }-\frac{\cosh \left(2 \pi  \sqrt{\frac{c \Delta_Q }{6}}\right)}{2 \sqrt{2} \pi \Delta_Q^{3/2}}\right]
\end{split}
\end{equation}
where $\Delta_Q:= (\Delta-c/24-Q^2/2)$.
One can also verify the above result in $\Delta\to\infty$ limit by doing the saddle point analysis, where the saddle of the relevant integral is given by $
    \beta = \pi \sqrt{\frac{c}{6\Delta_Q}}$.

To translate the above to the language of entanglement spectra, we define the following function $P_0:\mathbb{R}_+\to\mathbb{C}$: 
\begin{equation}\label{eeu1:main}
P_0(x):=|\widetilde{\bra{a}}0\rangle|^2 \left[\pi^2\left(\frac{c}{3}\right)^{3/2} \left(\frac{\sinh \left(x\sqrt{1-\frac{c\pi^2 Q^2}{3x^2}}\right)}{\left(x^2-\frac{\pi^2 c Q^2}{3}\right) }-\frac{\cosh \left(x\sqrt{1-\frac{c\pi^2 Q^2}{3x^2}}\right)}{\left(x^2-\frac{\pi^2 c Q^2}{3}\right)^{3/2}}\right)\right].
\end{equation}

The coarse-grained $P_Q$ and $P^{(N)}_Q$  (let us use the label $Q$ instead of $r$ to make explicit the reference to charge, the label of the irrep of $U(1)$) is given by 
\begin{equation}\label{eeu1:main2}
    P^{(N)}_Q(\lambda^{(N)})\sim P_0(y)\,,\quad P_Q(\lambda)\sim  P_0(x)
\end{equation}
where we define 
$y=2\sqrt{b\log(\lambda^{(N)}_{0,Q}/\lambda)}$ and $x=2\sqrt{b\log(\lambda_{0}/\lambda)}$
with $b:=\frac{c}{6}\log(\ell/\varepsilon)$. 
It is worth mentioning that here $a\sim b$ means $\lim a/b=1$. Finally, by integrating $P^{(N)}(\lambda)$ or $P(\lambda)$, we obtain the following formulas
\begin{equation}
\begin{aligned}\label{eeu1:main3}
     &n^{(N)}(\lambda,Q):=\int_{\lambda}^{\lambda^{(N)}_{0,Q}}d\lambda'\ P_Q^{(N)}(\lambda') \sim  n_0\left(2\sqrt{b\log(\lambda^{(N)}_{0,Q}/\lambda)}\right)\,,\\
     &n(\lambda,Q):=\int_{\lambda}^{\lambda_{0}}d\lambda'\ P_Q(\lambda') \sim  n_0\left(2\sqrt{b\log(\lambda_{0}/\lambda)}\right)\,,
\end{aligned} 
\end{equation}
where we have
\begin{equation}\label{eq:nlambda}
n_0(x):=\left(\frac{c}{3}\right)^{1/2} \frac{\exp\left(x\sqrt{1-\frac{c\pi^2 Q^2}{3x^2}}\right)}{2\sqrt{\left(x^2-\frac{\pi^2 c Q^2}{3}\right)} }.
\end{equation}

Before delving into proving the rigorous version of the above statements, we make two important remarks about eq.~\eqref{eeu1:main}
\begin{remark}
    a) For compact Lie group, our formula in eq.~\eqref{eeu1:main} is an improved version of what appears in \cite{Goldstein2018}. Furthermore, this is an explicit form compared to the integral form appearing in \cite{Goldstein2018}.

    b) Note that in large $x$ limit, the $\sinh$ piece in eq.~\eqref{eeu1:main} goes like $e^{x - \#\frac{Q^2}{x}}$.  The presence of $e^{-\#\frac{Q^2}{x}}$ involving the Casimir of the irrep in the case of compact Lie group has also been pointed out in \cite{Kang:2022orq} in the context of symmetry-resolved asymptotic density of states for CFT on spatial $S^{d-1}$ in any dimension.
    \end{remark}

  \subsection{Entanglement spectra and Numerical crosscheck I: }\label{app:spectrum}
As a $U(1)$ symmetric theory, we consider a free compact boson whose action we write for the first time in the manuscript here and it reads
\begin{equation}\label{eq:action1}
    \mathcal{S}=\frac{1}{8\pi}\displaystyle \int \mathrm{d}x_0\mathrm{d}x_1\partial_{\mu}\varphi\partial^{\mu}\varphi,
\end{equation}
where the target space of the real field $\varphi$ is compactified on a circle of radius $R$, i.e. $\varphi\sim \varphi +2\pi m R$ with $m \in \mathbb{Z}$. The compactification radius $R$ can be tuned such that our CFT predictions can be checked through numerical simulations for noninteracting fermions on a lattice (in terms of the Luttinger parameter $K$, this corresponds to taking $K=1$).
The spectrum of $\rho_A$ for free fermions
with $U(1)$ global charge can be determined numerically by using the eigenvalues $\nu_j$ of the two-point correlation matrix $C_{ij}$, which is a
$\ell\times \ell$ matrix provided by 
\begin{equation}
   C_{ij}=\frac{\sin[\pi(i-j)/2]}{\pi(i-j)}. 
\end{equation}

The $2^{\ell}$ eigenvalues $\lambda_j$ of the reduced density matrix are given by
\begin{equation}\label{eq:spectrumnj}
   \lambda_{n_j}= \prod_{j=1}^{\ell}\nu_j^{n_j}(1-\nu_j)^{1-n_j}
\end{equation}
where $\{n_j\}$ is a set of occupation numbers. Since for large $\ell$, the memory of an ordinary computer cannot store the $2^{\ell}$
eigenvalues $\lambda_{n_j}$,
we restrict to the eigenvalues with the largest value
and compute them with the following approximation. We
truncate the spectrum of the correlation matrix to the
first $M$ eigenvalues $\nu_j$ that maximize the distance with
respect to 0 and 1 and we compute $\lambda_{n_j}$ according to
eq.~\eqref{eq:spectrumnj}, with $j=1,\dots, M$.  The eigenvalues $\nu_j$ around 0 or 1 would produce a multiplicative factor that is near
to either 0 or 1 and, since we focus on the largest
eigenvalues of $\rho_A$, the multiplicative factor
we are missing should be close to 1.

If we are interested in a given charge sector, we need to properly choose the combination of the occupation numbers. For example, $Q=0$ corresponds to the half-filling case, where only half of the sites are occupied, and so on. Moreover, for each charge sector, we can normalize the spectrum dividing it by $\mathrm{Tr}\rho_{A,Q}$, depending on whether we analyze $\rho_{A,Q}$ or $\rho^{(N)}_{A,Q}$.

\begin{figure}[!ht]
     \centering
     \includegraphics[width=0.695\textwidth]{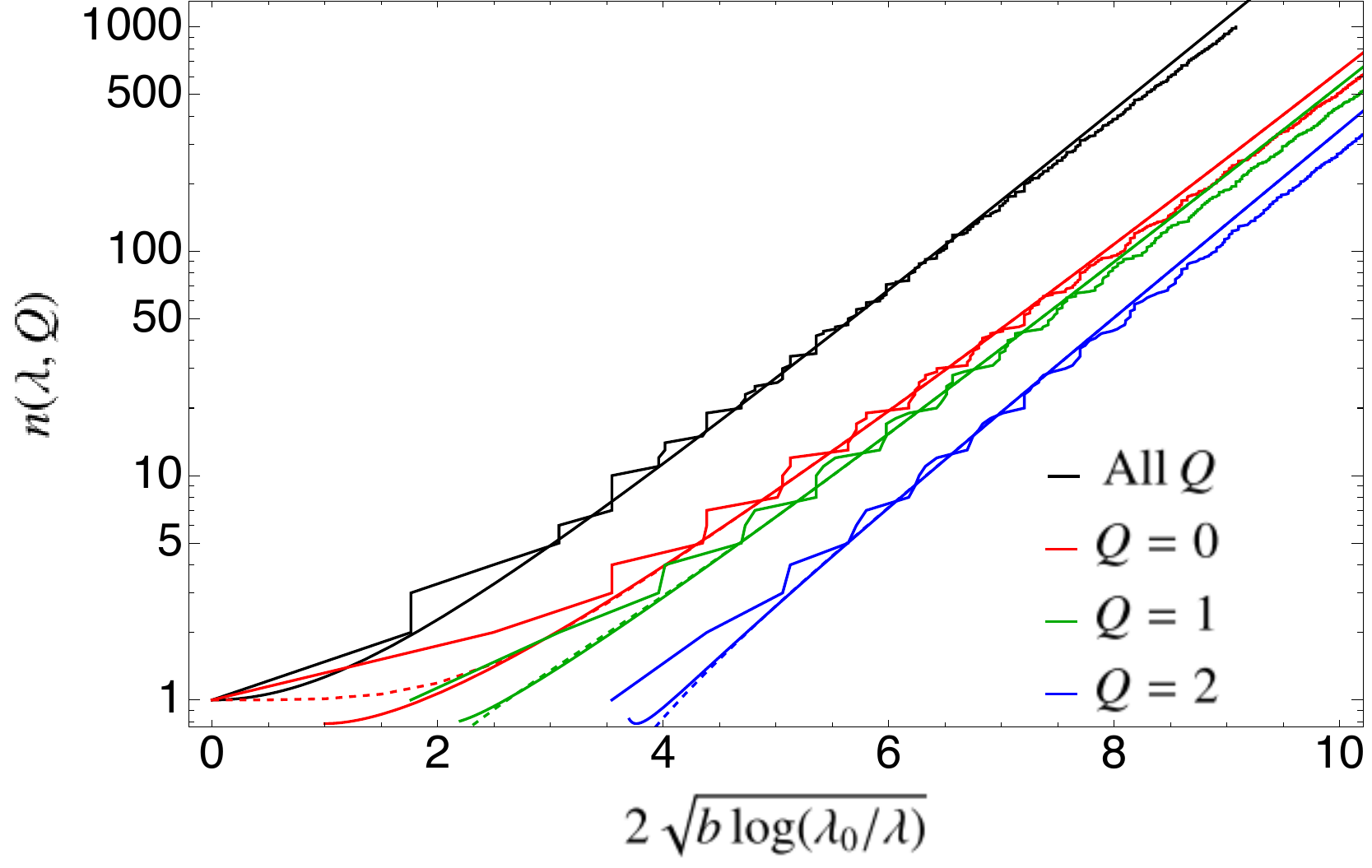}
     \caption{Density of the (non-normalized) entanglement
spectrum $n(\lambda, Q)$ for a subsystem of size $\ell=1000$ and $M=24$. The polygonal lines are the numerical points while the solid continue line is eq.~\eqref{eq:nlambda} in a given charge sector $Q$, and the total one is $n(\lambda)=I_0(x)$, where $x=2\sqrt{b\log(\lambda_{0}/\lambda)}$. This plot shows the matching between our result and the one in Ref. \cite{Goldstein2018} (where the authors used $\ell=10000$). Their prediction corresponds to the dashed lines, which completely overlaps with eq.~\eqref{eq:nlambda} as $x$ increases.}
     \label{fig:GS}
\end{figure}

So far, the only analytical result for the symmetry-resolved spectrum with a global $U(1)$ charge has been derived in \cite{Goldstein2018} for the (not-normalized) $\rho_{A,Q}$. In Fig. \ref{fig:GS}, we show that our result in eq.~\eqref{eq:nlambda} (full lines) coincide with the result of Ref. \ref{fig:GS} (dashed lines) in the limit of large $x$, where the scaling variable is $x=2\sqrt{b\log(\lambda_{0,Q=0}/\lambda)}$.  
Here we note that $\lambda_{0,Q=0}=\lambda_0$ and $b=-\log(\lambda_0)$ \cite{Lefevre_2008}.

Before ending this subsection, we also comment about the overlap $|\widetilde{\bra{a}}0\rangle|$ which enters eq.~\eqref{eeu1:main}. We have two possible boundary conditions (bc's) we can choose, Neumann (N, $\partial_{x_1}\varphi=0$) or Dirichlet (D, $\partial_{x_0}\varphi=0$) and we are interested in the case in which $Q$ is an integer number. In the presence of NN bc's, this amounts to choosing $R=2$, while for DD bc's, $R=1$ \cite{DiGiulio2022}. In the D case, the value of the bosonic field $\varphi$ may take different values at the
two boundaries of the annulus, while in the case of N bc's, the
structure is identical when describing the model in terms of the dual bosonic field $\theta$. Because of this duality, the partition function and the spectrum of states are invariant under $R\leftrightarrow 2/R$ \cite{DiFrancesco1997}. Using the results in \cite{DiGiulio2022}, for NN bc's we can compute $|\widetilde{\bra{a}}0\rangle|=\sqrt{R/2}=1$, while for DD we find $|\widetilde{\bra{a}}0\rangle|=1/\sqrt{R}=1$. This means that our results are compatible both with NN and DD bc's at the endpoint of our interval $A$. Despite the connection between the numerical entanglement spectrum and the boundary
conditions introduced around the entangling points by the regularisation procedure had been studied for a large class of models (for example, see \cite{lauchli2013operator,Alba:2017bgn,Di_Giulio_2020}), an analysis for the free fermionic case has not been done yet. Therefore, the matching between numerics and our analytical prediction represents the first analysis in this direction.

\subsection{Tauberian bounds and Numerical crosscheck II:}
In this subsection, we study 
\begin{equation}
    \int_{\lambda e^{-\beta\delta}}^{\lambda e^{\beta\delta}}d\lambda'\ P_Q(\lambda')\,,\quad \quad \& \quad \quad \int_{\lambda e^{-\beta\delta}}^{\lambda e^{\beta\delta}}d\lambda'\ P^{(N)}_Q(\lambda')\,,\quad\beta\equiv\frac{\pi^2}{\log(\ell/\varepsilon)},
\end{equation}
and put rigorous bounds on these quantities using Tauberian theorems, a toolkit from analytical number theory. In the following subsection, we prove these rigorous results. \\

For the finite group, we establish the following result:
\begin{theorem}[Tauberian theorem for EE spectra:finite group]\label{TBCFT2}
We consider the reduced density matrix $\rho$ corresponding to a single interval of length $\ell$, of a $1+1$D CFT with finite group symmetry $G$. The entanglement spectra projected onto irrep $r$, denoted as $P_r$, obeys the following inequality for $\beta=\frac{\pi^2}{\log(\ell/\varepsilon)}\to 0$:
    \begin{equation}\label{EEspec}
\begin{split}
& \frac{d^2_r}{|G|} \left(2\delta- 1\right)P_0(\lambda)\lesssim \int_{\lambda e^{-\beta\delta}}^{\lambda e^{\beta\delta}}d\lambda'\ P_r(\lambda')\lesssim  \frac{d^2_r}{|G|} \left(2\delta+ 1\right)P_0(\lambda)\,,\\
& \frac{d^2_r}{|G|} \left(2\delta- 1\right)P^{(N)}_0(\lambda)\lesssim \int_{\lambda e^{-\beta\delta}}^{\lambda e^{\beta\delta}}d\lambda'\ P^{(N)}_r(\lambda')\lesssim  \frac{d^2_r}{|G|} \left(2\delta+ 1\right)P^{(N)}_0(\lambda)\,,
 \end{split}
\end{equation}
where
\begin{equation}
\begin{split}
  P^{(N)}_0(\lambda)&:= |\widetilde{\bra{a}}0\rangle|^2  \left(\frac{c^4}{96\left(\frac{6b}{\pi^2 }\log\left(\frac{\lambda^{(N)}_{0,r}}{\lambda}\right)\right)^3}\right)^{1/4}\exp\left[2\sqrt{b\log\left(\frac{\lambda^{(N)}_{0,r}}{\lambda}\right)}\right]\,.\\
  P^{(N)}_0(\lambda)&:= |\widetilde{\bra{a}}0\rangle|^2  \left(\frac{c^4}{96\left(\frac{6b}{\pi^2 }\log\left(\frac{\lambda_{0,r}}{\lambda}\right)\right)^3}\right)^{1/4}\exp\left[2\sqrt{b\log\left(\frac{\lambda_{0}}{\lambda}\right)}\right]\,.
  \end{split}
\end{equation}
Here $b\equiv\frac{c}{6}\log(\ell/\varepsilon) $, $\varepsilon$ is the UV cut-off, $d_r$ is the dimension of the irrep $r$ and $|G|$ is the order of the group $G$. Furthermore, $\lambda^{(N)}_{0,Q}$ and $\lambda_{0,Q}$ are the maximum eigenvalue in charge $Q$ sector after and before normalizing the projected RDM, as defined around eq.~\eqref{eq:normEig}.
\end{theorem}

For continuous compact Lie group, we have the following result:
\begin{theorem}[Tauberian theorem for EE spectra: $U(1)$]\label{TBCFT3}
We consider the reduced density matrix $\rho$ corresponding to a single interval of length $\ell$, of a $1+1$D CFT with $U(1)$ symmetry. The entanglement spectrum projected onto charge $Q$, denoted as $P_Q$ (or $P_Q^{(N)}$) obeys the following inequality for $\beta=\frac{\pi^2}{\log(\ell/\varepsilon)}\to 0$:
    \begin{equation}\label{EEspecu1}
\begin{split}
&  \left(2\delta- 1\right) P^{u(1)}_{Q,0}(\lambda)\lesssim \int_{\lambda e^{-\beta\delta}}^{\lambda e^{\beta\delta}}d\lambda'\ P_Q(\lambda')\lesssim  \left(2\delta+ 1\right)|P^{u(1)}_{Q,0}(\lambda)\,,\\
& \left(2\delta- 1\right)P^{u(1),(N)}_{Q,0}(\lambda)\lesssim \int_{\lambda e^{-\beta\delta}}^{\lambda e^{\beta\delta}}d\lambda'\ P^{(N)}_Q(\lambda')\lesssim  \left(2\delta+ 1\right)P^{u(1),(N)}_{Q,0}(\lambda)
 \end{split}
\end{equation}
where
\begin{equation}\label{eeu1:mainrg}
\begin{split}
  P^{u(1)}_{Q,0}(\lambda)&:= \widetilde{\bra{a}}0\rangle|^2 \left(\frac{c^3}{48\left(\frac{6b}{\pi^2 }\log\left(\frac{\lambda_{0,Q}}{\lambda}\right)\right)^2}\right)^{1/2}\exp\left(f(\lambda)-\frac{\pi^2\sqrt{c}}{6f(\lambda)}Q^2\right)\,,\\
  P^{u(1),(N)}_{Q,0}(\lambda)&:=\widetilde{\bra{a}}0\rangle|^2 \left(\frac{c^3}{48\left(\frac{6b}{\pi^2 }\log\left(\frac{\lambda_{0,Q}}{\lambda}\right)\right)^2}\right)^{1/2}\exp\left(f^{(N)}(\lambda)-\frac{\pi^2\sqrt{c}}{6f^{(N)}(\lambda)}Q^2\right)\,
  \end{split}
\end{equation}
and
\begin{equation}
    f(\lambda):=2\sqrt{b\log\left(\frac{\lambda_{0}}{\lambda}\right)} \,,\quad f^{(N)}(\lambda):=2\sqrt{b\log\left(\frac{\lambda^{(N)}_{0,Q}}{\lambda}\right)}\,.
\end{equation}
Here $b\equiv\frac{c}{6}\log(\ell/\varepsilon) $, $\varepsilon$ is the UV cut-off, $\lambda^{(N)}_{0,Q}$ and $\lambda_{0,Q}$ are the maximum eigenvalue in charge $Q$ sector after and before normalizing the projected RDM, as defined around eq.~\eqref{eq:normEig}. Here $a\lesssim b$ means $\lim a/b \leqslant 1$.
\end{theorem}
\begin{remark}
    The naive answer given by eq.~\!\eqref{eeu1:main} in large $x$ or $y$ limit (where $y=2\sqrt{b\log(\lambda^{(N)}_{0,Q}/\lambda)}$ and $x=2\sqrt{b\log(\lambda_{0}/\lambda)}$) reproduces the above asymptotic form given in eq.~\!\eqref{eeu1:mainrg}. In particular, one can replace the above lower and upper bound by $P^{\pm}(f(\lambda))$ or $P^{\pm}(f^{(N)}(\lambda))$, where we have
    \begin{equation}\label{eq:f1pm}
    P^{\pm}(w)= |\widetilde{\bra{a}}0\rangle|^2\frac{(2\delta\pm 1)\pi^2 c^{3/2}}{3^{3/2} }\left(\frac{\sinh \left(w\sqrt{1-\frac{\pi^2 Q^2}{3w^2}}\right)}{\left(w^2-\frac{\pi^2  Q^2}{3}\right) }-\frac{\cosh \left(w\sqrt{1-\frac{\pi^2 Q^2}{3w^2}}\right)}{\left(w^2-\frac{\pi^2 Q^2}{3}\right)^{3/2}}\right)\,.
\end{equation}
\end{remark}

\begin{remark}
Noting that the lower bound is positive as soon as $2\delta>1$, we derive that there has to be at least one eigenvalue in the interval $(\lambda e^{-\beta\delta},\lambda e^{\beta\delta})$. Hence we obtain the following result about spacing of $\lambda_{i,r}$ away from the continuum limit (in the limit $\lambda_{i,r}\to 0$)
\begin{equation}
\log\frac{\lambda_{i,r}}{\lambda_{i+1,r}} \leqslant \frac{\pi^2}{\log(\ell/\varepsilon)}\,, \quad \ell/\varepsilon\to\infty\,, \quad \lambda_{i+1,r}<\lambda_{i,r}\to 0\,.
\end{equation}
\end{remark} 

We now present a numerical crosscheck of the above bounds for the continuous Lie group.\\

\begin{figure}[t]
     \centering
     \includegraphics[width=0.515\textwidth]{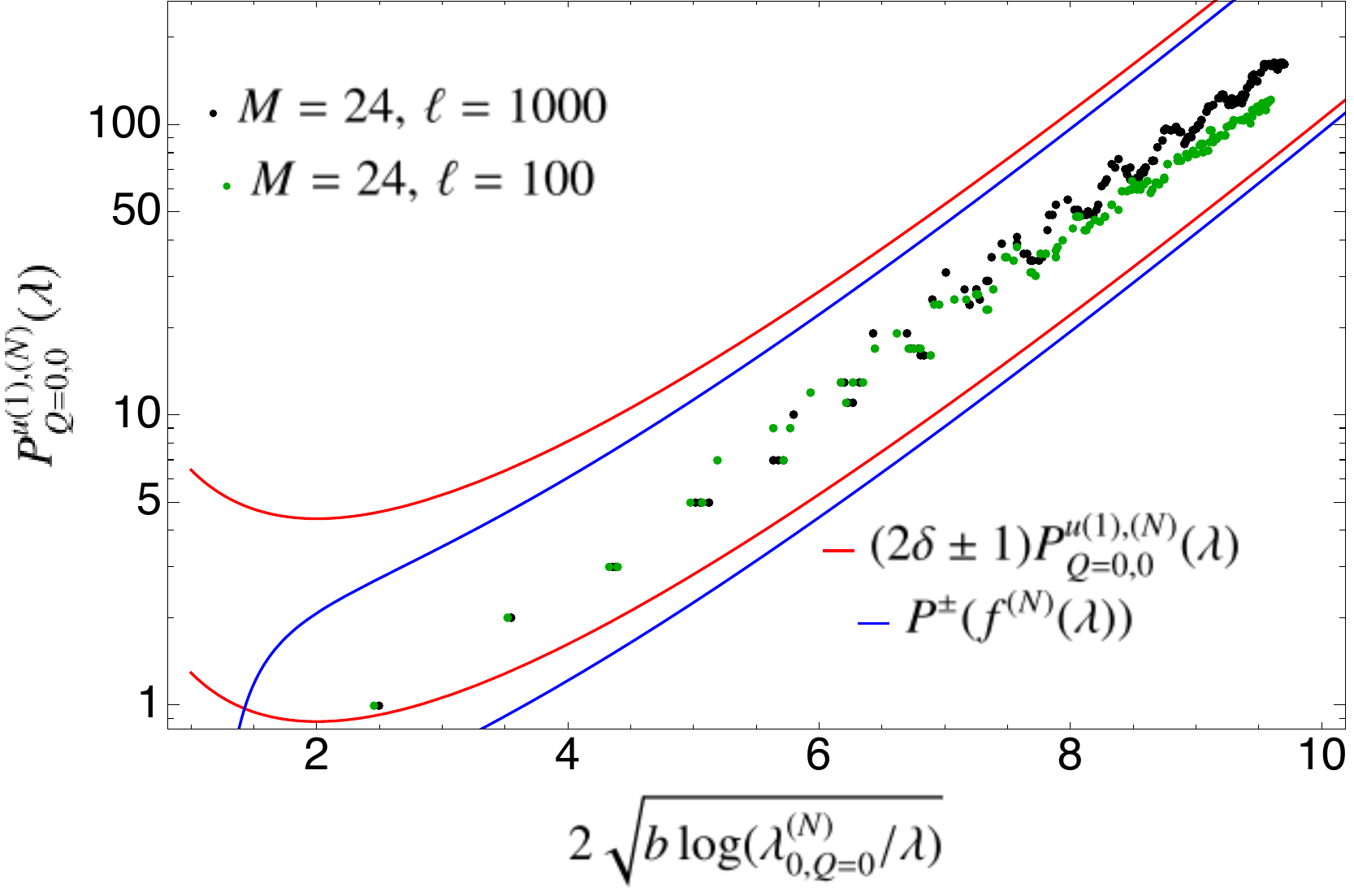}
     \includegraphics[width=0.495\textwidth]{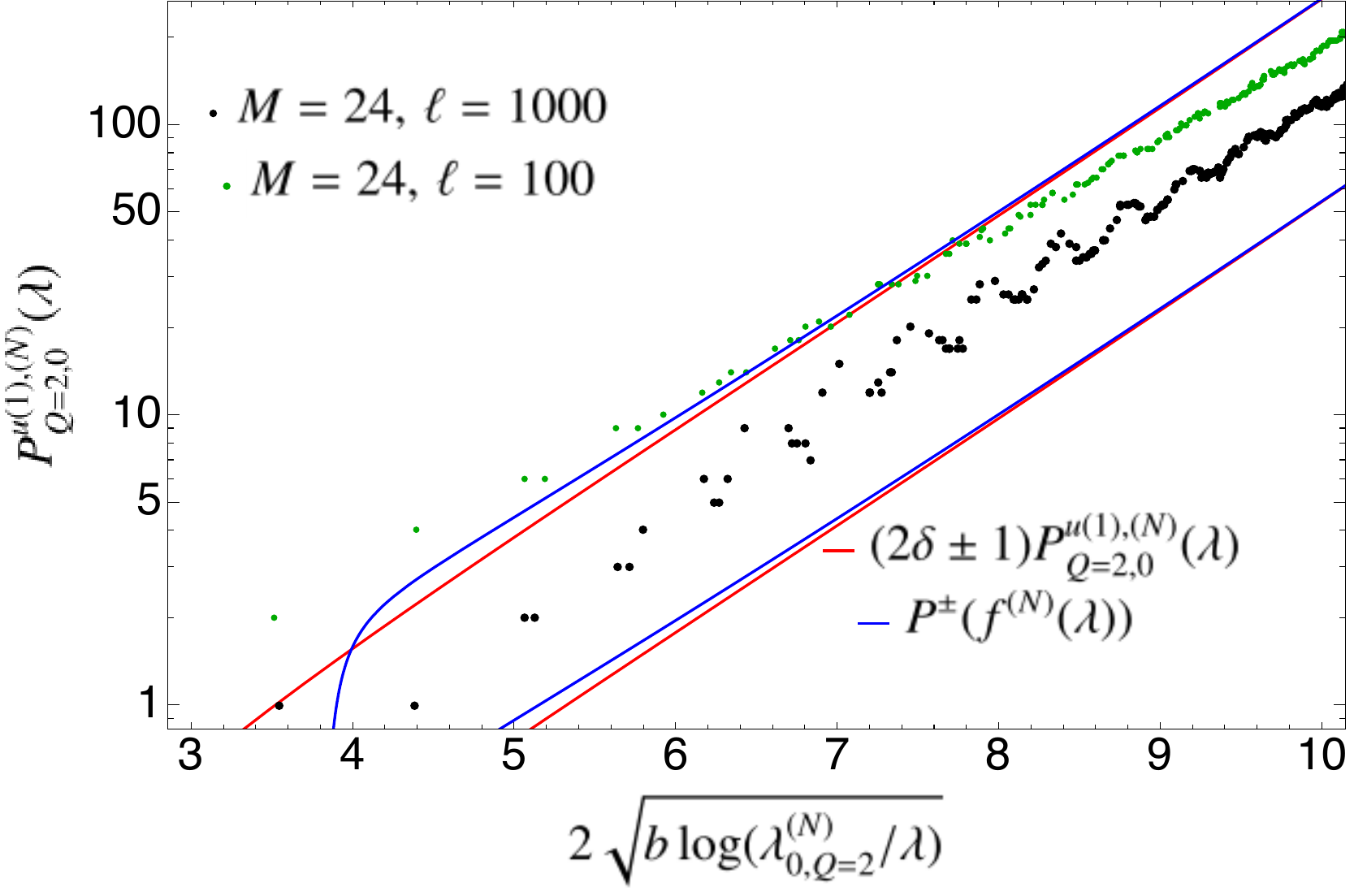}
    \includegraphics[width=0.495\textwidth]{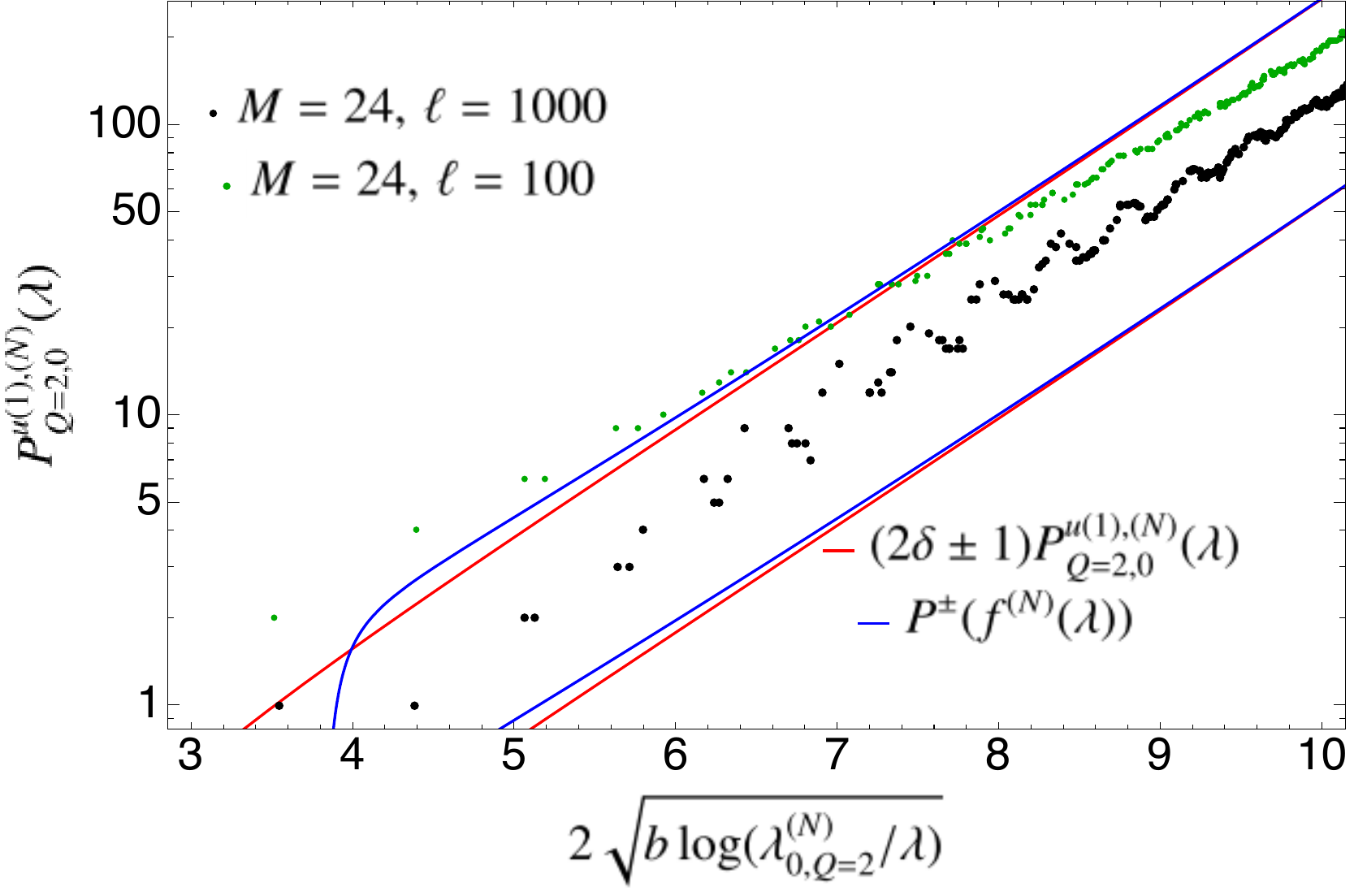}
     \caption{Numerical check of eq.~\eqref{EEspecu1} for a free fermionic system with conserved particle number. The geometry is an interval of size $\ell$ on the infinite line. The symbols are the numerical points obtained using the procedure described in Section \ref{app:spectrum} to evaluate the (normalized) entanglement spectrum in each charge sector $Q$. The solid red lines correspond to $(2\delta \pm 1) \pi ^2 e^{x-\frac{\pi ^2Q^2}{6 x}} \sqrt{\frac{1}{108 x^4}}$ with $\delta=0.75$ and $x=2\sqrt{b\log(\lambda_{0,Q}/\lambda)}$, while the blue ones correspond to eq.~\eqref{eq:f1pm} with $c=1$.}
     \label{fig:spectrum}
\end{figure}
In order to cross-check eq.~\eqref{EEspecu1}, we use again the numerical techniques of Section \ref{app:spectrum} for noninteracting fermions, which are equivalent to a free compact boson with $K=1$.

We can compute the eigenvalues $\lambda$ in a given charge sector $Q$ and we set $b=-\log \lambda_{0}$, where $\lambda_{0}$ is the maximal eigenvalue of $\rho_A$ \cite{Lefevre_2008}, while $\lambda^{(N)}_{0,Q}=\lambda_0/\mathrm{Tr \rho_{A,Q}}$, according to eq.~\eqref{eq:normEig}. 

\begin{figure}[t]
     \centering
     \includegraphics[width=0.495\textwidth]{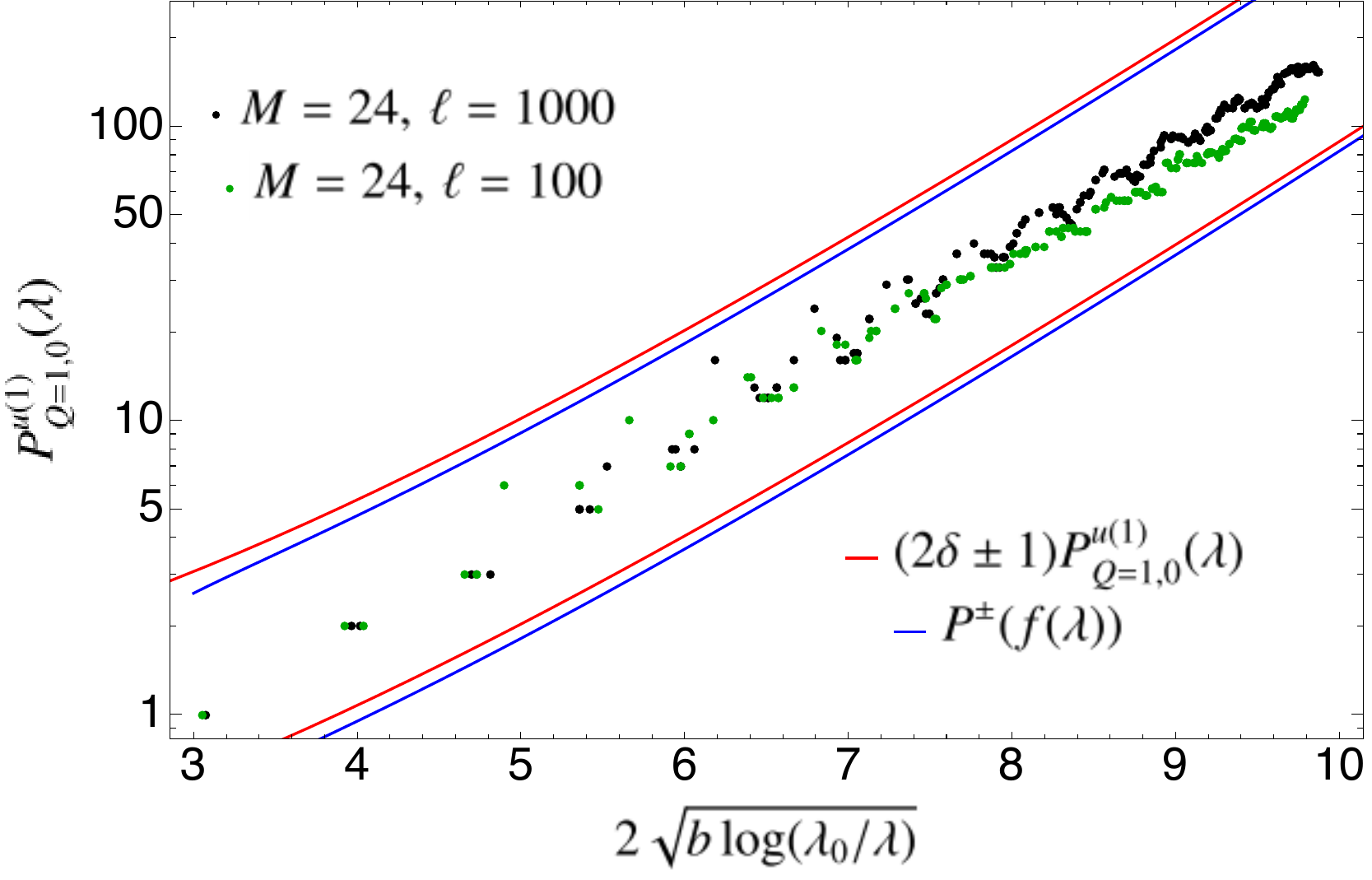}
    \includegraphics[width=0.495\textwidth]{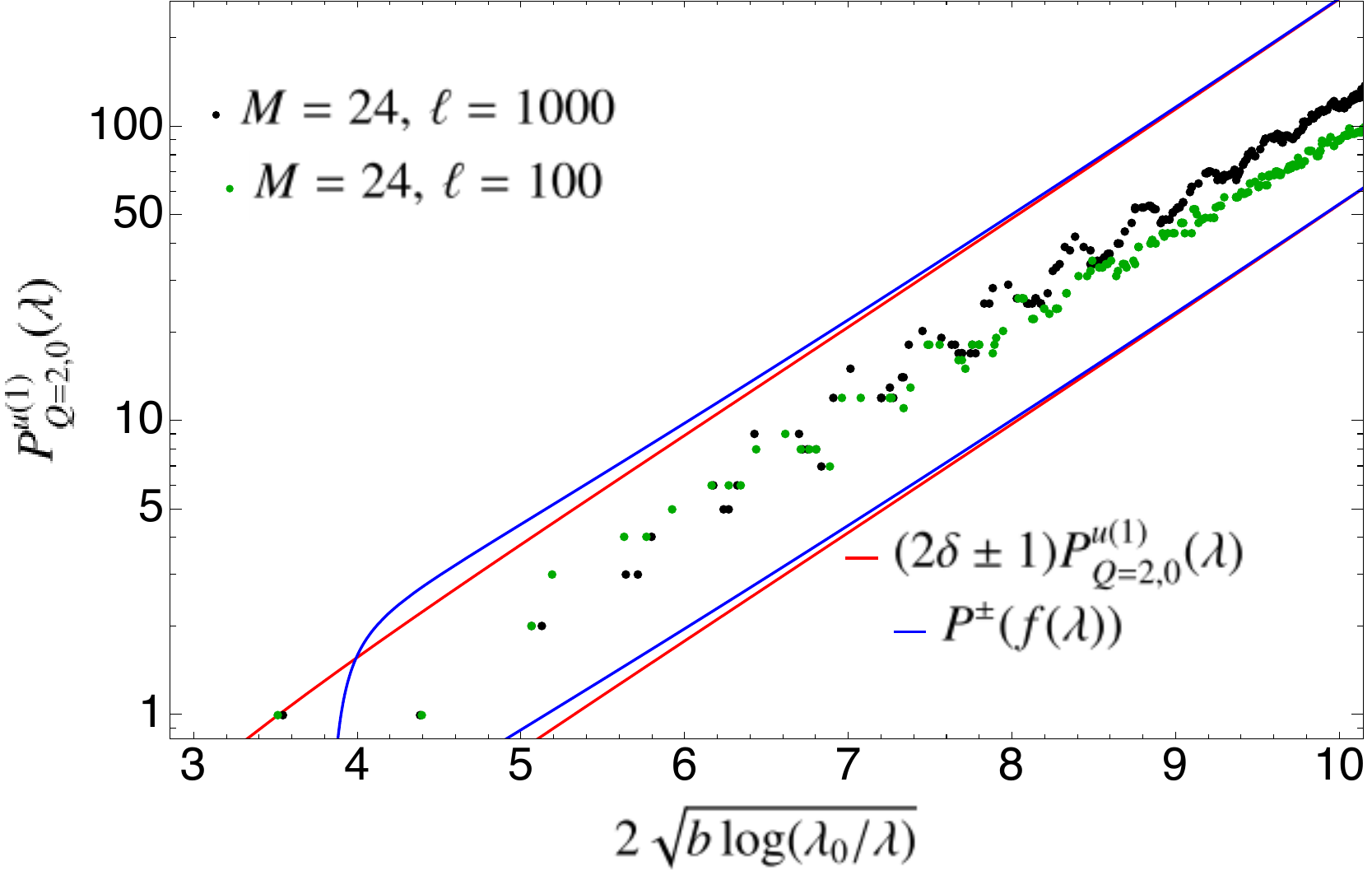}
     \caption{Same numerical check as in Fig. \ref{fig:spectrum} but now the entanglement spectrum in each charge sector is not normalized (i.e. $\sum_i\lambda_{i,Q}<1$)
     and the scaling variable is $x=2\sqrt{b\log(\lambda_0/\lambda)}$, as done in Ref. \cite{Goldstein2018}. }
     \label{fig:spectrumGS}
\end{figure}

We count the number of eigenvalues in the interval $[e^{-\beta \delta}\lambda,e^{\beta \delta}\lambda]$,  where $\beta=\pi^2/(6b)$, and we plot them as a function of $x=2\sqrt{b\log(\lambda^{(N)}_{0,Q}/\lambda)}$ in Fig. \ref{fig:spectrum}, using $|\langle \widetilde{a}|0\rangle |^2=1$, normalizing the spectrum in each charge sector. The symbols correspond to the numerical values computed using the techniques described in the previous subsection \ref{app:spectrum}. We notice that, as $Q$ increases, the agreement between our predictions and the exact lattice results requires bigger subsystem sizes $\ell$. This is due to the fact that our formulas hold for the largest eigenvalues of  $\rho_A$, which fall in the charge sectors closest to $Q=0$. Therefore, in order to see a better agreement, we need to increase both $\ell$ (the subsystem size of $A$) and $M$ (the truncation of the spectrum --- see Section \ref{app:spectrum} for more details about its derivation). If we do not normalize $\rho_{A,Q}$, the result does not change for $Q=0$, while for $Q=1$ and $Q=2$ the matching between numerics and analytics improves already for $\ell=100$, as we show in Fig. \ref{fig:spectrumGS}. We also comment that in all the figures, we plot both eq.~\eqref{EEspecu1} (red line) and~\eqref{eq:f1pm}
(blue line). Note the former can be derived from the latter in the limit of large $x$, i.e. $\lambda$ much smaller than $\lambda_{0}$. \\

Now we come back to the Tauberian analysis. The theorem~\ref{TBCFT2} and \ref{TBCFT3} follow from similar theorems stated in the language of BCFT owing to the eqs.~\!\eqref{spectrumBCFT1},~\eqref{spectrumBCFT2},~\eqref{eq:normEig}. In what follows, we will go through this Tauberian analysis in BCFT.

\subsection{Tauberian Analysis in BCFT}
In the BCFT approach to entanglement entropy, the entanglement spectra i.e. eigenvalues of the modular Hamiltonian are captured by the open string partition function. Hence knowing details about entanglement spectra amounts to knowing the density of states in the open string channel. The $\epsilon\to 0$ limit of the partition function provides us with asymptotic density of states, which, in turn, can be leveraged to derive universal scaling law of entanglement spectra. Note the statement in terms of entanglement entropy is a statement about the partition function i.e. a statement in canonical ensemble. Thus symmetry resolution in entanglement spectra amounts to symmetry resolution in microcanonical ensemble. The technical challenge is to derive a microcanonical version of the symmetry resolution from canonical version. In the naive approach, this is achieved via inverse Laplace transformation. However, a precise way of symmetry resolution in the microcanonical version requires a meticulous approach since the density of states is a distribution for a unitary compact CFT with discrete spectrum of operators/states.

\subsubsection{Tauberian framework}
Before delving into the details of the problem in hand, let us zoom out and explain the general philosophy and scope of the Tauberian analysis. We recall that the proper way to treat distributions is to integrate them against test functions from appropriate space of functions. Ideally, we would want to integrate them over a characteristic function of a small energy interval and estimate them:

\begin{equation}
  \label{teqN}
   \int_{\Delta-\delta}^{\Delta+\delta}d\Delta'\ \mu(\Delta')\,, 
\end{equation}
where $\mu$ stands for a non-negative measure.
For our purpose, $\mu$ can be the full density of states or the density of states projected onto a particular irrep of the group $G$.

The following obvious inequality turns out to be useful to connect the above quantity defined in the microcanonical ensemble to the canonical one:
\begin{equation}\label{teq0}
  e^{\beta(\Delta-\delta)}\int_{\Delta-\delta}^{\Delta+\delta}d\Delta'\ \mu(\Delta')e^{-\beta\Delta'}\leqslant  \int_{\Delta-\delta}^{\Delta+\delta}d\Delta'\ \mu(\Delta') \leqslant e^{\beta(\Delta+\delta)}\int_{\Delta-\delta}^{\Delta+\delta}d\Delta'\ \mu(\Delta')e^{-\beta\Delta'}\, ,
\end{equation}
By taking the limit $\beta\to 0$ with the upper and lower bounds approaching each other, the integral appearing in the lower and the upper bound gets dominant contribution from $\Delta\to\infty$ regime. We observe that
\begin{equation}\label{teq}
   e^{\beta(\Delta\pm\delta)}\int_{\Delta-\delta}^{\Delta+\delta}d\Delta'\ \mu(\Delta')e^{-\beta\Delta'}= 2\delta e^{\beta(\Delta\pm\delta)} \int_{-\infty}^{\infty} dt\ Y(\beta+it)e^{i\Delta t} \frac{\sin(\delta t)}{\delta t},
\end{equation}
where $Y(\beta):=\int d\Delta' \mu(\Delta')e^{-\beta\Delta}$ is an expression in the canonical ensemble. For our purpose, as we will see later, $Y(\beta)=e^{-\beta c/24} K(\beta)$ and $K(\beta)$ can be the full partition function $Z(\beta)$ or the projected one i.e $Z_r(\beta)$.

The integral in eq.~\!\eqref{teq} goes over the entire range of $t$ from $-\infty$ to $\infty$. However, for sufficiently large values of  $t$, $Y(\beta+it)$ recurs and it becomes harder to estimate the integral. 
To address the aforementioned issue, it becomes necessary to restrict the $t$ integral inside $(-\Lambda,\Lambda)$ regime for some $\Lambda=O(1)$. This allows us to avoid the recurrence and control the integral. However, such a restriction in the $t$ integral can potentially be unrelated to the estimate of $\mu$ integrated against some nicely behaved function. In other words, we would still like to make a statement about eq.~\eqref{teqN}. To achieve this we approximate the characteristic function of the interval $(-\delta,\delta)$ by bandlimited functions $\Phi_{\pm}(x)$ i.e 
\begin{equation}\label{eq:basicIneq}
\Phi_-(x)\leqslant \Theta(x\in(-\delta,\delta))\leqslant \Theta(x\in[-\delta,\delta])\leqslant \Phi_+(x)\,,\quad \text{supp}[\widehat{\Phi}_{\pm}] \subset (-\Lambda,\Lambda)\,,
\end{equation}

Now observe the following:
For $-\delta\leqslant x$, $e^{-\beta(x+\delta)}\leqslant 1$ and for $x\leqslant \delta $, $e^{-\beta(x-\delta)}\geqslant 1$, thus for $-\delta\leqslant x\leqslant \delta $, $e^{-\beta(x+\delta)}\leqslant 1$ and  $e^{-\beta(x-\delta)}\geqslant 1$. Also note that, when $x\notin[-\delta,\delta]$ we trivially have $e^{-\beta(x+\delta)}\Theta(x\in(-\delta,\delta]))=\Theta(x\in(-\delta,\delta))=0$  and $e^{-\beta(x-\delta)}\Theta(x\in[-\delta,\delta])=\Theta(x\in[-\delta,\delta])=0$. Therefore we obtain for all $x$,
\begin{equation}\label{anotheIneq}
   e^{-\beta(x+\delta)}\Theta(x\in(-\delta,\delta]))\leqslant \Theta(x\in(-\delta,\delta))\leqslant \Theta(x\in[-\delta,\delta])\leqslant e^{-\beta(x-\delta)} \Theta(x\in[-\delta,\delta])\,.
\end{equation}
Now we combine eqs.~\eqref{anotheIneq} and \eqref{eq:basicIneq} to derive 
\begin{equation}\label{eq:NbasicIneq}
\begin{split}
&e^{-\beta(x+\delta)}\Phi_-(x)\leqslant e^{-\beta(x+\delta)}\Theta(x\in(-\delta,\delta]))\leqslant \Theta(x\in(-\delta,\delta))\leqslant \Theta(x\in[-\delta,\delta])\\
&\leqslant e^{-\beta(x-\delta)}\Theta(x\in[-\delta,\delta])
\leqslant e^{-\beta(x-\delta)}\Phi_+(x)\,,
\end{split}
\end{equation}

At this point, we substitute $x=\Delta'-\Delta$ in eq.~\eqref{eq:NbasicIneq} and integrate against non-negative measure $\mu(\Delta')$ to obtain
\begin{equation}
\begin{split}
 & e^{\beta(\Delta-\delta)} \int_{0}^{\infty}d\Delta'\ \mu(\Delta')\Phi_-(\Delta'-\Delta)e^{-\beta\Delta'} \\
  &\leqslant \int_{0}^{\infty }d\Delta'\ \mu(\Delta') \Theta(\Delta'-\Delta\in(-\delta,\delta))\\
   &\leqslant \int_{0}^{\infty }d\Delta'\ \mu(\Delta') \Theta(\Delta'-\Delta\in[-\delta,\delta])\\
  &\leqslant   e^{\beta(\Delta+\delta)}\int_{0}^{\infty}d\Delta'\ \mu(\Delta')\Phi_+(\Delta'-\Delta)e^{-\beta\Delta'}\,.
   \end{split}
\end{equation}

Now it is easier to estimate the lower and the upper bound:$$ e^{\beta(\Delta\pm\delta)} \int_{\Delta-\delta}^{\Delta+\delta}d\Delta'\ \mu(\Delta')\Phi_{\pm}(\Delta'-\Delta)e^{-\beta\Delta}=e^{\beta(\Delta\pm\delta)}\int_{-\Lambda}^{\Lambda} dt\ Y(\beta+it)e^{-i\Delta t} \widehat{\Phi}_{\pm}(t)\,.$$

With the above general picture in mind, we come back to the problem at hand i.e. we will analyze the entanglement spectra by using a more rigorous treatment using Tauberian theory, applied to the framework of BCFT. We will obtain an asymptotic result with \textbf{optimal} error estimates.

\subsubsection{BCFT with finite group symmetry}
\begin{theorem}[Tauberian Theorem in BCFT: finite group]\label{TBCFT}
    We consider a BCFT with finite group symmetry $G$. Let us define $\rho_r$ to be the density of states, transforming under irrep $r$ of $G$, in the open string Hilbert space with $G$-invariant boundary state $\widetilde{\ket{a}}$. We have
    \begin{equation}
    |\widetilde{\bra{a}}0\rangle|^2\frac{d_r^2}{|G|}(2\delta-1)\rho_*(\Delta)\lesssim\int_{\Delta-\delta}^{\Delta+\delta}d\Delta'\  \rho_r(\Delta') \lesssim|\widetilde{\bra{a}}0\rangle|^2\frac{d_r^2}{|G|}(2\delta+1)\rho_*(\Delta)\,,
\end{equation}
where $$\rho_*(\Delta):=\left(\frac{c}{96\Delta^3}\right)^{1/4}\exp\left[2\pi\sqrt{\frac{c\Delta}{6}}\right]\,.$$ and by $a\lesssim b$ we mean
    \begin{equation}
    	\begin{split}
    		\lim\frac{a}{b}\leqslant1.
    	\end{split}
    \end{equation}
\end{theorem} 

\begin{remark}
   a) From the analysis of \cite{Mukhametzhanov2020}, it can be shown that the bounds are optimal when the window width $2\delta$ is an integer. 
   
   b) When $2\delta$ is not an integer, while the constants $2\delta\pm 1$ provide correct bounds, they can be replaced by some other function of $\delta$, leading to a tighter bound. For more details, see section.~\!$5$ of \cite{Mukhametzhanov2020} and the references therein, in particular \cite{littmann2013quadrature}.
\end{remark}

The rest of the subsection is devoted to establishing the above theorem. We recall that the open string partition function projected onto $r$th irrep is denoted as $Z_r(\beta)$ and defined by
\begin{equation}
Z_r(\beta):= \int_{0}^{\infty}d\Delta'\  \rho_r(\Delta')e^{-\beta(\Delta-c/24)}\,.
\end{equation}

In the general framework presented above, we should identify $\mu=\rho_r$ and $Y(\beta)=e^{-\beta c/24}Z_r(\beta)$. This leads to 
\begin{equation}
    I_{-;r}(\Delta)\leqslant \int_{\Delta-\delta}^{\Delta+\delta}d\Delta'\  \rho_r(\Delta')\leqslant I_{+;r}(\Delta)
\end{equation}
where 
\begin{equation}
I_{\pm;r}(\Delta):= e^{\beta(\Delta\pm\delta-c/24)} \int_{-\Lambda}^{\Lambda}dt\ Z_r(\beta+it) \widehat{\Phi}_{\pm}(t) e^{-itc/24},
\end{equation}

Leveraging the character orthogonality theorem, we have 
\begin{equation}\label{ZrZg}
Z_r(\beta)= \frac{d_r}{|G|} \sum_{g\in G} \chi_r^*(g) Z(g,\beta)\,.
\end{equation}
We use eq.~\eqref{ZrZg} to replace $Z_r(\beta+it) $ in the expression for $I_{\pm;r}(\Delta)$, then exchange the sum and integral to obtain 
\begin{equation}
I_{\pm;r}(\Delta)=\frac{d_r}{|G|} \sum_{g\in G} \chi_r^*(g) e^{\beta(\Delta\pm\delta-c/24)} \int_{-\Lambda}^{\Lambda}dt\ Z(g,\beta+it) \widehat{\Phi}_{\pm}(t) e^{-itc/24}.
\end{equation}
At this point, we use the modular transformation to go to the closed string channel:
\begin{equation}
I_{\pm;r}(\Delta)=\frac{d_r}{|G|} \sum_{g\in G} \chi_r^*(g) e^{\beta(\Delta\pm\delta-c/24)} \int_{-\Lambda}^{\Lambda}dt\ Z_g\left(\frac{4\pi^2}{\beta+it}\right) \widehat{\Phi}_{\pm}(t) e^{-itc/24}.
\end{equation}
Here $Z_g\left(\frac{4\pi^2}{\beta+it}\right) $ is the amplitude in closed string channel with the presence of defect (extended along the finite width).

Now we take $Z_g\left(\frac{4\pi^2}{\beta+it}\right) $ and divide this into light part, denoted as $Z_{g;L}$ and heavy part, denoted as $Z_{g;H}$:
\begin{equation}
\begin{split}
Z_{g;L}\left(\frac{4\pi^2}{\beta+it}\right)&:=\sum_{\Delta_g<c/24}  |  \widetilde{\langle a|}  \Delta_g\rangle|^2\ e^{-\frac{4\pi^2}{\beta+it}(\Delta_g-c/24)}, \\
Z_{g;H}\left(\frac{4\pi^2}{\beta+it}\right)&:=\sum_{\Delta_g\geqslant c/24}  |  \widetilde{\langle a|}  \Delta_g\rangle|^2 \ e^{-\frac{4\pi^2}{\beta+it}(\Delta_g-c/24)},
\end{split}
\end{equation}
where $ \widetilde{\ket{a}} $  is the boundary state. The idea is to show that with appropriate choice of $\Lambda$, the contribution to the integral from $Z_{g;H}\left(\frac{4\pi^2}{\beta+it}\right)$ is suppressed compared to that from $Z_{g;L}\left(\frac{4\pi^2}{\beta+it}\right)$ and to establish that $Z_{g;L}\left(\frac{4\pi^2}{\beta+it}\right)$ gives the leading answer that we seek for. 

In order to achieve the above, let us define
\begin{equation}\label{def:I}
I_{\pm;r;i}(\Delta):=\frac{d_r}{|G|} \sum_{g\in G} \chi_r^*(g) e^{\beta(\Delta\pm\delta-c/24)} \int_{-\Lambda}^{\Lambda}dt\ Z_{g;i}\left(\frac{4\pi^2}{\beta+it}\right) \widehat{\Phi}_{\pm}(t) e^{-itc/24}\,, \ i=L,H\,.
\end{equation}
and observe the simple inequality
\begin{equation}
 I_{\pm;r;L}(\Delta)-| I_{\pm;r;H}(\Delta) |\leqslant I_{\pm;r}(\Delta)=I_{\pm;r;L}(\Delta)+I_{\pm;r;H}(\Delta) \leqslant I_{\pm;r;L}(\Delta)+| I_{\pm;r;H}(\Delta) |\,.
\end{equation}
Now we will estimate $I_{\pm;r;L}(\Delta)$ and $I_{\pm;r;H}(\Delta)$ separately.  Clearly, we have
\begin{equation}\label{eq:heavy}
\begin{split}
| I_{\pm;r;H}(\Delta)|&\leqslant \frac{d_r}{|G|} \sum_{g\in G} d_r e^{\beta(\Delta\pm\delta-c/24)} \int_{-\Lambda}^{\Lambda}dt\ \bigg|Z_{g;H}\left(\frac{4\pi^2}{\beta+it}\right)\bigg| \big|\widehat{\Phi}_{\pm}(t)\big| \\
&=\frac{d^2_r M}{|G|} \sum_{g\in G} e^{\beta(\Delta\pm\delta-c/24)} \int_{-\Lambda}^{\Lambda}dt\ \bigg|Z_{g;H}\left(\frac{4\pi^2}{\beta+it}\right)\bigg| \\
&\leqslant \frac{d^2_r M}{|G|} e^{\beta(\Delta\pm\delta-c/24)}\sum_{g\in G} \int_{-\Lambda}^{\Lambda}dt\ Z_{g;H}\left(\frac{4\pi^2\beta}{\beta^2+t^2}\right)\\
&\leqslant \frac{2d^2_r M\Lambda}{|G|} e^{\beta(\Delta\pm\delta-c/24)} \sum_{g\in G}Z_{g;H}\left(\frac{4\pi^2\beta}{\beta^2+\Lambda^2}\right),
\end{split}
\end{equation}
where in the first line we used $|\chi_r(g)|\leqslant d_r$, in the second line we used $\text{Max}\big|\widehat{\phi}_{\pm}(t)\big|=M$ is some order one number.  In the third line, we used 
$$\bigg|\sum_{x\geqslant 0} g(x)^2 e^{-\left(\frac{4\pi^2}{\beta+it}\right)x}\bigg|\leqslant  \sum_{x\geqslant 0} g(x)^2e^{-\left(\frac{4\pi^2\beta}{\beta^2+t^2}\right)x}\,.$$ 
In the fourth line we used
$$\sum_{x\geqslant 0} g(x)^2e^{-\left(\frac{4\pi^2\beta}{\beta^2+t^2}\right)x}\leqslant \sum_{x\geqslant 0} g(x)^2e^{-\left(\frac{4\pi^2\beta}{\beta^2+\Lambda^2}\right)x}\,, |t|\leqslant \Lambda\  $$
and integrate over $t$ to get an extra factor of $2\Lambda$.
We have
\begin{equation}
Z_{g;H}\left(\frac{4\pi^2\beta}{\beta^2+\Lambda^2}\right)\leqslant Z_{g}\left(\frac{4\pi^2\beta}{\beta^2+\Lambda^2}\right)=Z\left(g, \frac{\beta^2+\Lambda^2}{\beta}\right).
\end{equation}
Here in the first step, we bound the heavy part by the full partition function and then in the second step we use modular invariance to go to open string channel. 
In $\beta\to 0$ limit, $\left(\frac{\beta^2+\Lambda^2}{\beta}\right)\to\infty$, thus $Z\left(g, \frac{\beta^2+\Lambda^2}{\beta}\right)$ is dominated by the vacuum, which we assume to be invariant under $G$ action. As a result, we obtain
\begin{equation}
Z_{g;H}\left(\frac{4\pi^2\beta}{\beta^2+\Lambda^2}\right)\leqslant Z\left(g, \frac{\beta^2+\Lambda^2}{\beta}\right)\sim e^{\frac{\Lambda^2}{\beta}\frac{c}{24}}.
\end{equation}

In the $\beta\to 0$ limit we have 
\begin{equation}
Z_{g;H}\left(\frac{4\pi^2\beta}{\beta^2+\Lambda^2}\right)\leqslant Z_{g}\left(\frac{4\pi^2\beta}{\beta^2+\Lambda^2}\right)=Z\left(g, \frac{\beta^2+\Lambda^2}{\beta}\right)\sim e^{\frac{\Lambda^2}{\beta}\frac{c}{24}}.
\end{equation}

Plugging this estimate of $Z_{g;H}\left(\frac{4\pi^2\beta}{\beta^2+\Lambda^2}\right)$  in eq.~\eqref{eq:heavy} we obtain in $\beta\to 0$ limit
\begin{equation}\label{eq:heavyestimate}
| I_{\pm;r;H}(\Delta)|\leqslant 2d^2_r M\Lambda e^{\beta(\Delta\pm\delta-c/24)\frac{\Lambda^2}{\beta}\frac{c}{24}}.
\end{equation}
Now we come back to the estimation of the light part. For a finite central charge the light part gets contribution from finite number of states with $\Delta<c/24$ and the sum over group elements is also finite, so we can do the integrals term by term in the saddle point approximation in $\beta\to0$ limit. In particular, we want to set\footnote{With this choice of $\beta$, the saddle point analysis can be made completely rigorous using dominated convergence theorem as done in appendix A of \cite{Pal:2023cgk}.} $\beta=\pi\sqrt{\frac{c}{6\Delta}}$ and take $\Delta\to \infty$. With this choice, we obtain
\begin{equation}\label{eq:lightestimate}
\begin{split}
I_{\pm;r;L}(\Delta)&\sim  \frac{d^2_r}{|G|}e^{\beta(\Delta\pm\delta-c/24)} \frac{1}{2\pi}\left(2\delta\pm \frac{2\pi}{\Lambda}\right) |\widetilde{\bra{a}}0\rangle|^2\sqrt{\frac{6\beta^3}{\pi c}}\exp\left[\frac{4\pi^2}{\beta}\frac{c}{24}\right],
\end{split}
\end{equation}

The contribution from the excited states of each twisted sector is exponentially suppressed compared to the ground state of that twisted sector. We combine this fact with the inequality $\Delta_{g0} >\Delta_{e0}$ (here $\Delta_{g0}$ refers to the scaling dimension of the operator corresponding to the ground state of the defect Hilbert space, twisted by $g$) for $g\neq e$ to deduce that the $g=e$ term gives the leading contribution in the sum appearing the in the expression for $I_{\pm;r;L}(\Delta)$.  Furthermore, we used the bandlimited function from \cite{Mukhametzhanov2020} such that $2\pi\widehat{\Phi}_{\pm}(0)=2\delta\pm \frac{2\pi}{\Lambda}$.

Now we compare eq.~\eqref{eq:heavyestimate} and eq.~\eqref{eq:lightestimate}; we choose $\Lambda<2\pi$ (it can be arbitrarily close to $2\pi$) to show that the contribution from the heavy part is subleading\footnote{By more careful estimate, one can set $\Lambda=2\pi$ and prove that the heavy part is polynomially suppressed \cite{Mukhametzhanov2020}}. Hence,
upon making the choice $\beta=\pi\sqrt{\frac{c}{6\Delta}}$ and $\Lambda=2\pi$ in eq.~\eqref{eq:lightestimate}, we obtain the desired upper and the lower bound, that  
concludes the proof of theorem~\ref{TBCFT}.

\subsubsection{Extension to compact Lie group}
The generalization to compact Lie group involves using the character orthogonality theorem appropriate for the compact Lie group. The analogue of eq.~\eqref{ZrZg} reads
\begin{equation}\label{ZrZgC}
Z_r(\beta)= \int_{G} dg\ \chi_r^*(g) Z(g,\beta)\,.
\end{equation}
Here $dg$ is the unit-normalized Haar-measure of the group $G$. 

Let us explain the analysis for $U(1)$, which can suitably be generalized with little extra effort. For $U(1)$, the irreps are labelled by integer charge $Q$ and the group elements are given by $g(\theta)=e^{2\pi i\theta}$. Thus we have
\begin{equation}\label{ZrZgU1}
Z_Q(\beta)= \int_{-1/2}^{1/2}d\theta\ e^{-2\pi i Q\theta}Z(\theta,\beta)\,.
\end{equation}

The eq.~\eqref{def:I} will read
\begin{equation}\label{def:I2}
I_{\pm;r;i}(\Delta):=\int_{-1/2}^{1/2}d\theta\ e^{-2\pi i Q\theta} e^{\beta(\Delta\pm\delta-c/24)} \int_{-\Lambda}^{\Lambda}dt\ Z_{\theta;i}\left(\frac{4\pi^2}{\beta+it}\right) \widehat{\Phi}_{\pm}(t) e^{-itc/24}\,, \ i=L,H\,.
\end{equation}

The heavy part can be estimated in same way since $|\int d\theta e^{-2\pi iQ\theta} f(\theta)| \leqslant\int d\theta |f(\theta)| $. A similar analysis has been performed in \cite{Pal:2020wwd} for the torus partition function. 

The estimation of the light part requires knowing the ground state energy of twisted Hilbert space corresponding to the group element $g=e^{2\pi i\theta}$. We have
\begin{equation}
    Z_{\theta;L}\left(\frac{4\pi^2}{\beta+it}\right)\sim \widetilde{\langle a|}  \Delta_0(\theta)\rangle|^2  e^{\frac{4\pi^2}{\beta+it}(c/24-\theta^2/2)}.
\end{equation}
Here $|\Delta_0(\theta)\rangle$ is the ground state in the twisted Hilbert space with scaling dimension $\Delta_0(\theta)$. 
If we plug this in the expression for $I_{\pm;r;L}(\Delta)$ and exchange the $\theta$ integral with $t$ integral
\begin{equation*}\label{Gauss}
I_{\pm;r;L}(\Delta):=e^{\beta(\Delta\pm\delta-c/24)}  \int_{-\Lambda}^{\Lambda}dt\ \widehat{\Phi}_{\pm}(t) e^{-itc/24} e^{\frac{4\pi^2}{\beta+it}c/24}\int_{-1/2}^{1/2}d\theta\ \widetilde{\langle a|}  \Delta_0(\theta)\rangle|^2 e^{-2\pi i Q\theta} e^{-\frac{4\pi^2\theta^2}{2(\beta+it)}}\,,
\end{equation*}

Assuming $ \widetilde{\langle a|}  \Delta_0(\theta)\rangle|^2 $ is a slow varying function, one can do a saddle point approximation for the $\theta$ integral. The saddle is given by $\theta=0$ and consequently, we have 
\begin{equation}
    I_{\pm;r;L}(\Delta) \sim e^{\beta(\Delta\pm\delta-c/24-\frac{Q^2}{2})}  \int_{-\Lambda}^{\Lambda}dt\ \widehat{\Phi}_{\pm}(t) e^{-it(c/24+Q^2/2)} e^{\frac{4\pi^2}{\beta+it}c/24} \sqrt{\frac{\beta+it}{2\pi}}\,.
\end{equation}

Now the remaining integral in $t$ variable is dominated by $t=0$ point. This can be seen in an elementary way by focusing on the main term $e^{\frac{4\pi^2}{\beta+it}c/24}$ and realizing that
\begin{equation*}
    \bigg|e^{\frac{4\pi^2}{\beta+it}c/24}\bigg|=e^{\frac{4\pi^2\beta}{\beta^2+t^2}c/24} = \exp\left[\frac{4\pi^2\beta}{\beta^2}c/24-\frac{\pi ^2 c}{6 \beta^3}t^2+\cdots \right]\,.
\end{equation*}

Hence, we have
\begin{equation*}
\begin{split}
    I_{\pm;r;L}(\Delta)&\sim \widehat{\Phi}_{\pm}(0)e^{\beta(\Delta\pm\delta-c/24)} e^{\frac{4\pi^2}{\beta}c/24-\frac{\beta Q^2}{2}} \sqrt{\frac{\beta}{2\pi}} \int_{-\infty}^{\infty} dt\ \exp\left[-\frac{\pi ^2 c}{6 \beta^3}t^2+\cdots \right]\\
 &\sim  \widehat{\Phi}_{\pm}(0)e^{\beta(\Delta\pm\delta-c/24)} e^{\frac{4\pi^2}{\beta}c/24-\frac{\beta Q^2}{2}} \sqrt{\frac{\beta}{2\pi}} \left(\beta^{3/2} \sqrt{\frac{6}{c\pi}}\right).
\end{split}
\end{equation*}

Now just like the analysis for the finite group, we make sure $I_{\pm;r;H}(\Delta)$ is suppressed by choosing $\Lambda<2\pi$ and setting $\beta=\pi\sqrt{c/6\Delta}$. Finally, when the dust settles, upon setting, $2\pi\widehat{\Phi}_{\pm}(0)=2\delta\pm 1$ we obtain

\begin{theorem}[Tauberian Theorem in BCFT: $U(1)$]\label{TBCFTU1}
    We consider a BCFT with $U(1)$ symmetry. Let us define $\rho_Q$ to be the density of states with charge $Q$ in the open string Hilbert space with $U(1)$-invariant boundary state $\widetilde{\ket{a}}$. We have
    \begin{equation}
    |\widetilde{\bra{a}}0\rangle|^2(2\delta-1)e^{-\pi\sqrt{\tfrac{c}{24\Delta}} Q^2}\rho^{u(1)}_*(\Delta)\lesssim\int_{\Delta-\delta}^{\Delta+\delta}d\Delta'\  \rho_r(\Delta') \lesssim|\widetilde{\bra{a}}0\rangle|^2(2\delta+1)e^{-\pi\sqrt{\tfrac{c}{24\Delta}}  Q^2}\rho^{u(1)}_*(\Delta)\,,
\end{equation}
where $$\rho^{u(1)}_*(\Delta):=\left(\frac{c}{48\Delta^2}\right)^{1/2}\exp\left[2\pi\sqrt{\frac{c\Delta}{6}}\right]\,.$$ and by $a\lesssim b$ we mean
    \begin{equation}
    	\begin{split}
    		\lim\frac{a}{b}\leqslant1
    	\end{split}
    \end{equation}
\end{theorem} 
\begin{remark}
    a) Compared to the theorem.~\!\ref{TBCFT}, here the polynomial $\Delta^{\#}$ in the bound is different. 
    
    b) We denote the presence of $e^{-\pi\sqrt{\tfrac{c}{24\Delta}}  Q^2}$, which depends on the Casimir of the irrep. In $\Delta\to \infty$ limit, this term goes to $1$. A similar term has also been pointed out in \cite{Kang:2022orq} in the context of symmetry-resolved asymptotic density of states for CFT on spatial $S^{d-1}$ in any dimension.
\end{remark}

\section{Conclusions}\label{sec:conclusions}

In this paper, we study the symmetry resolution of the entanglement of a CFT with a given global symmetry $G$. For this purpose, we rely on the BCFT description of our system and we choose symmetry-preserving boundary conditions. This approach allows us to derive the leading contributions in  $\ell/\varepsilon$ to the charged moments and to the SREE for \textit{any} finite and compact Lie group $G$. This result confirms the equipartition of the entanglement already found for continuous symmetries \cite{sierra,calabrese2021symmetryresolved} and it shows that the first $O(1)$ correction which spoils it, depends on the dimension of the irreducible representation. 

Beyond the interesting findings about the physics of symmetry resolution, we provide here for the first time an explicit connection between the charged moments, which are cornerstones in computing the SREE, and the topological defects. Indeed, we can explicitly construct the Verlinde line generating the symmetry and computing its action on the (symmetric) boundary states of our theory, which basically gives the charged moments of the reduced density matrix. In order to support our findings, we provide some examples to which we apply our general scheme. 
In addition to this, we also establish a connection between SREE and partition function of an orbifold CFT, which allows us to bypass the computation of the charged moments. In this setup, the SREE is related to a closed string amplitude with a proper boundary state in the orbifold CFT. 
In the last part of this paper, we study how the entanglement spectrum of the reduced density matrix behaves in a given charge sector of a CFT invariant under a finite or compact Lie group $G$. We study in detail the $U(1)$ case, comparing our result with the ones of Ref. \cite{Goldstein2018}, and we derive optimal and rigorous upper and lower bounds on the integrated symmetry-resolved spectrum. 

We suggest some remaining questions and interesting future directions.

It would be worth understanding how the formalism developed here about the connection between the charged moments and the topological defects can be modified when $G$ is not a global symmetry of our system, and therefore this equivalence does not hold anymore. This would allow us to study within a field theory setup the recently introduced entanglement asymmetry, i.e. a subsystem measure of the symmetry breaking \cite{Ares_2023}. 
Another interesting direction is to understand the symmetry resolution of non-abelian groups for other entanglement measures, like the negativity, which has been studied only in the abelian case so far \cite{Cornfeld_2018,Murciano_2021}. The main challenge is that an equivalent BCFT description for the partial transpose reduced density matrix is not available so far. 

By using the dimensional reduction techniques \cite{Casini_2007}, one might be able to study the symmetry resolution in higher dimensional systems, since it would allow for a reduction to a $1+1$ dimensional problem, where we can exploit the results we found in this manuscript. 
Another approach toward higher dimensions would be to use the thermal effective field theory as in \cite{Kang:2022orq, Benjamin:2023qsc}.
An analysis has been done only for the free massive scalar theory across a hyperplane in
generic dimensions (see \cite{Murciano_2020}), and it would be interesting to address this problem for a CFT with an arbitrary symmetry. 
In this direction, the holography can be a helpful tool.  In fact, in the context of holography, holographic charged Renyi entropy in any dimensions has been investigated in \cite{Belin:2013uta}, building up on \cite{Hung:2011nu} and \cite{Casini:2011kv}.

Another interesting future work would be studying the symmetry resolution of entanglement entropy with respect to non-invertible symmetries. The symmetry resolution of a torus partition function has been done for non-invertible symmetries in \cite{Lin:2022dhv}. One may be able to extend this to the symmetry-resolved entanglement entropy. Further exciting avenues include considering the symmetry resolution of various measures of multipartite entanglement such as the ones introduced in \cite{Gadde:2023zzj} \cite{Gadde:2022cqi}. 

\section*{Acknowledgments}
We thank Filiberto Ares, Pasquale Calabrese, Giuseppe Di Giulio, Michele Fossati, Kantaro Ohmori, Brandon Rayhaun, Shu-Heng Shao, Yuji Tachikawa, and Yijian Zou for useful discussions and comments on the draft.
The work by YK, HO, and SP is supported in part by the U.S. Department of Energy, Office of Science, Office of High Energy Physics, under Award Number DE-SC0011632. In addition, YK is supported by the Brinson Prize Fellowship at Caltech.
HO is supported in part by the Simons Investigator Award (MP-SIP-00005259), the World Premier International Research Center Initiative, MEXT, Japan, and
JSPS Grants-in-Aid for Scientific Research 20K03965 and 23K03379. 
SP is supported in part by the Sherman Fairchild Postdoctoral Fellowship at Caltech.
SM thanks support from Caltech Institute for Quantum Information and Matter and the Walter Burke Institute for Theoretical Physics at Caltech. 
This work was performed in part at
the Aspen Center for Physics, which is supported by NSF grant PHY-1607611.

\appendix

%%%%%%%%%%%%%%%%%%%%%%%%%%%%%%%%%%%%%%%%%%%%%%%%%%%%%%%%%%%%%%%%%%%%%%%%%%%%%%%%%%%%%%%%%%%%%%
\section{$\omega$-twisted boundary state}\label{sec:twisted boundary}
%%%%%%%%%%%%%%%%%%%%%%%%%%%%%%%%%%%%%%%%%%%%%%%%%%%%%%%%%%%%%%%%%%%%%%%%%%%%%%%%%%%%%%%%%%%%%%

Let $G$ be a group of automorphisms of the chiral algebra $\ca{A}$ and $\ca{I}$ be a set of all irreducible representations of $\ca{A}$.
The twisted boundary condition by $\omega \in G$ is given by
\begin{equation}\label{eq:tw condition}
\pa{W_n - \pa{-1}^h \omega\pa{\ti{W}_{-n}}}\widetilde{\ket{b; \omega}}=0,
\end{equation}
where $W_n$ are the modes of the current $W(z)$ associated with the chiral algebra $\ca{A}$ and $\ti{W}_{-n}$ are those of the antiholomorphic chiral algebra $\ti{\ca{A}}$.
$h$ is the conformal dimension of $W_n$.
By using a unitary representation of $\omega \in G$, $R(\omega) : \ca{V}_\lambda \to \ca{V}_{\omega \pa{\lambda}} $,
we have
\begin{equation}
\omega(W_n) = R(\omega) W_n R^{-1}(\omega)
\end{equation}
and the solution to the twisted boundary condition can be given as a linear combination of the twisted Ishibashi states,
\begin{equation}
\kett{(\mu,\lambda^*) ; \omega}
\equiv
R(\omega) \kett{(\mu,\mu^*)}
=
\sum_N R(\omega) \ket{\mu;N} \otimes \overline{\ket{\mu;N}}.
\end{equation}
where $\kett{(\mu,\mu^*)}$ is the Ishibashi state and $\ket{\mu:N}$ means a state in the Verma module $\mu \in \ca{I}$ labeled by $N$.
The unitary operator $R(\omega)$ acts only on the anti-holomorphic part.
The state $(\mu,\lambda^*)$ is the highest-weight state of $\ca{A} \times \ti{\ca{A}}$ and satisfies $\lambda = \omega(\mu)$.
One can easily see that the twisted Ishibashi states satisfy the twisted boundary condition.
In the following, we abbreviate the twisted Ishibasi state as $\kett{\lambda;\omega}$.

In the same as the boundary state,
not all linear combinations of the twisted Ishibashi states can be the twisted boundary state.
It is constrained by the open-closed duality.
One can construct the twisted boundary state in a similar way to the Cardy construction of the Cardy boundary condition.
Let $\ca{A}^\omega$ be the twisted chiral algebra generated by the twisted current,
\begin{equation}
W(\ex{2\pi i }z) = \omega(W(z)),
\end{equation}
and $\ca{I}^\omega$ be a set of all irreducible representations of $\ca{A}^\omega$.
A twisted analog of some consistency conditions from the open-closed duality is given by
\begin{equation}\label{eq:twduality}
\widetilde{\bra{0; \text{id}}} \ti{q}^{\fr{1}{2} H_\text{closed}} \widetilde{\ket{ \lambda;  \omega }} = \chi_{\ti{\lambda}  } (\tau),
\end{equation}
where $\chi_{\ti{\lambda} }$ is the character of $\ti{\lambda} \in \ca{I}^\omega$ and $\ti{q} = \ex{- \fr{2\pi i}{\tau}}$.
To give the twisted Cardy boundary state, we first investigate the property of the character.
The modular $S$ transformation of the character $\chi_{\ti{\lambda} }$ can be spanned by the twining character,
\begin{equation}
\chi_\mu^\omega(\tau) = \tr_{\ca{H}_\mu} R(\omega) q^{L_0-\fr{c}{24}}.
\end{equation}
Note that $\chi_\mu^\omega(\tau) = 0$ if $\omega \notin \ca{S}(\mu)$ where $\ca{S}(\mu)$ is the stabilizer of $\mu$.
By using the twining character, the modular $S$ transformation of the character $\chi_{\ti{\lambda} }$ is given by
\begin{equation}
\chi_{\ti{\lambda}  } (\tau) =
\sum_{\mu \in \ca{I}(\omega)}
S^\omega_{\ti{\lambda} \mu} \chi^\omega_\mu \pa{-\fr{1}{\tau}},
\end{equation}
where $\ca{I}(\omega) = \{  \lambda \in \ca{I} | \omega(\lambda) = \lambda   \}$. Note that $\lambda \in \ca{I}(\omega) \Leftrightarrow \omega \in \ca{S}(\lambda)$.
Using the matrix $S^{\omega}_{\ti{\lambda} \mu}$, we can give the solution to the twisted Cardy condition (\ref{eq:twduality}) in a diagonal CFT as
\begin{equation}
\widetilde{\ket{\lambda; \omega}} = \sum_{\mu\in\ca{I}(\omega)} \fr{ S^{\omega}_{\ti{\lambda \mu}} }{\sqrt{S_{0\mu}}} \kett{\mu ; \omega}.
\end{equation}
We call this the $\omega$-twisted Cardy state.
We would like to emphasize that this $\omega$-twisting should be distinguished from the twisting in eq. \eqref{eq:twS}.

\clearpage
\bibliographystyle{JHEP}
\bibliography{main}

\end{document}